\documentclass{acm_proc_article-sp}

\usepackage{epsfig,url}
\usepackage{amssymb}
\usepackage{amsmath}
\usepackage{amsfonts}
\usepackage{graphicx}
\usepackage{color}
\usepackage{listing}
\usepackage{float}
\usepackage{amssymb}
\usepackage{epsfig}
\usepackage{color}
\usepackage{algorithm}
\usepackage{algpseudocode}
\usepackage{subfigure}
\usepackage{xspace}

\usepackage[pdfpagelabels,
           plainpages=false,
           bookmarks=true,
           hypertexnames=false,
           pdffitwindow=false,
           colorlinks=true,
           linkcolor=blue,
           citecolor=blue,
           filecolor=black,
           urlcolor=black,
           pdfauthor={},
           pdftitle={XXX},
           pdfsubject={2014},
           pdfkeywords={YYY},
           pdfproducer={Latex with hyperref},
           pdfcreator={pdflatex}
          ]{hyperref}

\long\def\comment#1{}

\definecolor{red}{rgb}{1,0,0}

\newcommand{\tool}{\textit{YouLighter}\xspace}
\newcommand{\toolintitle}{YOULIGHTER\xspace}
\newcommand{\distance}{\textit{Constellation Distance}\xspace}
\newcommand{\distintitle}{Constellation Distance\xspace}

\newcommand{\node}{{edge-node}\xspace}
\newcommand{\nodes}{{edge-nodes}\xspace}

\newcommand{\tstat}{{AcmeSniff}\xspace}

\newcommand{\TApub}{\textit{ISP1-A}\xspace}
\newcommand{\TApri}{\textit{ISP1-B}\xspace}

\newcommand{\TBpub}{\textit{ISP1-C}\xspace}

\newcommand{\TC}{\textit{ISP2}\xspace}

\newcommand{\AMS}{E-1\xspace} 
\newcommand{\FRA}{E-2\xspace} 
\newcommand{\MIL}{E-3\xspace} 
\newcommand{\SWI}{E-4\xspace} 
\newcommand{\SWII}{E-5\xspace} 
\newcommand{\LND}{E-6\xspace} 
\newcommand{\MAD}{E-7\xspace} 

\title{YouLighter: An Unsupervised Methodology\\
to Unveil YouTube CDN Changes}

\numberofauthors{9}

\author{
\alignauthor Danilo Giordano\\
       \affaddr{DET, Politecnico di Torino, Italy}\\
       \email{danilo.giordano@polito.it}
\alignauthor Stefano Traverso\\
       \affaddr{DET, Politecnico di Torino, Italy}\\
       \email{stefano.traverso@polito.it}
\alignauthor Luigi Grimaudo\\
       \affaddr{DAUIN, Politecnico di Torino, Italy}\\
       \email{luigi.grimaudo@polito.it} \and
\alignauthor Marco Mellia\\
       \affaddr{DET, Politecnico di Torino, Italy}\\
       \email{marco.mellia@polito.it}
\alignauthor Elena Baralis\\
       \affaddr{DAUIN, Politecnico di Torino, Italy}\\
       \email{elena.baralis@polito.it} \and 
\alignauthor \hspace{-0pt}Alok Tongaonkar\\
       \affaddr{\hspace{-0pt}Symantec, USA}\\
\email{\vspace{-2pt}alok\_tongaonkar@symantec.com} 
\alignauthor \textcolor{white}{.}\\
       \affaddr{\textcolor{white}{.}}\\
       \email{\textcolor{white}{.}}\\
\alignauthor  Sabyasachi Saha\\
       \affaddr{Symantec, USA}\\
\email{sabyasachi\_saha@symantec.com}
}

\begin{document}

\maketitle
\begin{abstract} 
YouTube relies on a massively distributed Content Delivery Network (CDN)
to stream the billions of videos in its catalogue.  Unfortunately,  very little information about the design of such CDN is available. This, combined with the pervasiveness of YouTube, poses a big challenge for Internet Service
Providers (ISPs), which are compelled to optimize end-users' Quality of Experience (QoE) while having no control on the CDN decisions.

This paper presents \tool, an
unsupervised technique to identify changes in the
YouTube CDN. \tool leverages only passive
measurements to cluster co-located identical caches
into {\em \nodes}. This automatically unveils the structure of YouTube's CDN. Further, we propose a new metric, called {\em \distance}, that compares the clustering obtained from two different time snapshots, to pinpoint sudden changes.
While several approaches allow comparison between the clustering results from the {\em same} dataset,
no technique allows to measure the similarity of clusters from {\em different} datasets.
Hence, we develop a novel methodology, based on the \distance, to solve this problem.

By running \tool over 10-month long traces obtained from two ISPs in different countries,
we pinpoint both sudden changes in \node allocation, and small alterations to the cache allocation policies which actually impair the QoE that the end-users perceive.
\end{abstract}

%
%

\section{Introduction}
\label{sec:intro}

YouTube is one of the most popular and demanding Internet Video services. It accounts for 1 billion users distributed world-wide, who watch 6 billion hours of videos per month.\footnote{\url{https://www.youtube.com/yt/press/statistics.html}}
Due to its popularity and the nature of the content that it distributes, it is very challenging for both Google to maintain the YouTube infrastructure, and for Internet Service Providers (ISPs) to manage the underlying network. They both have to optimize the video delivery to improve customers' Quality of Experience (QoE).

YouTube leverages a massive, globally distributed Content Delivery Network (CDN), the Google CDN~\cite{CalderIMC13}, to handle such a demanding load. Indeed, it consists of hundreds of {\it \nodes} scattered in the Internet. Each \node hosts hundreds of video servers, or {\it caches}, which can each potentially serve any video a user may request~\cite{Adhikari2012_infocom}. Once a user starts a video playback, the CDN load balancing algorithm directs the request to one of the caches. There is no known way to determine in advance which cache, or which \node will be used. And sadly, ISPs have no way to influence this choice~\cite{Torres_2011,Cas2014}.

YouTube CDN is in continuous evolution, and its proprietary and never disclosed design makes it one of the most challenging CDNs to monitor and measure. Unsurprisingly, this spurred a lot of interest in the research community which investigated and disclosed some of the design secrets behind YouTube infrastructure -- see Sec.~\ref{sec:related} for details.

While understanding YouTube CDN internals is interesting, it is even more challenging to design a system that monitors and automatically identifies changes in the CDN that could ultimately affect the QoE that the users perceive. Changes may involve modification in the infrastructure, e.g., the activation of a new \node, or in the load balancing algorithm decision, e.g., a sudden switch of caches to serve requests.

In this paper, we present \tool, a novel methodology to automatically monitor and pinpoint changes in the YouTube CDN. \tool relies on an unsupervised learning approach that, as such, does not require any knowledge of the YouTube infrastructure. Instead, it only assumes that the ISP has deployed passive probes, which expose TCP flow level logs summarizing video requests from users.
Considering a given observation window of, say one day, \tool aggregates these flow logs to constitute a {\it snapshot} of the traffic exchanged with YouTube servers.
Based on DBSCAN~\cite{ester1996density} clustering, which is a well-established unsupervised machine learning algorithm, \tool is able to automatically group thousands of caches into less than 10 \nodes using simple features that characterize the network distance of caches from the vantage point.

\tool runs the clustering algorithm as soon as a new snapshot is available. The challenge then is to compare the two results obtained considering two consecutive snapshots, i.e., compare two different datasets with the ultimate goal of highlighting changes in the CDN. Unfortunately, no standard methodology is available to compare clusters obtained from different datasets. Hence, we propose a generic framework that solves this problem. We transform each snapshot into a {\it constellation}, and we compare two constellations using the notion of \distance. The bigger the distance between two snapshots, the more different the sets of caches YouTube uses to serve ISP customers during the two corresponding periods of time.
\tool thus highlights changes in the \nodes used by YouTube to serve ISPs' customers. Moreover, \tool can also pinpoint deviations from the typical behavior of the YouTube \nodes, e.g., due to congestion arising in the network which makes the same \nodes look different. Thus, \tool has the potential to unveil sudden changes caused by the YouTube CDN infrastructure and unveil possible QoE issues for ISP customers.

We validate our methodology over different datasets we collect from the different vantage points that we have deployed in two ISPs in two different countries for 10 months.
First, we demonstrate that the clustering algorithm \tool adopts is effective at identifying and grouping YouTube caches belonging to different \nodes when considering a snapshot.
Second, and more importantly, we run \tool over different collected snapshots considering the longitudinal dataset. We pinpoint several examples of sudden and previously undiscovered changes in the YouTube CDN policies. For some of them, we drill down showing the impact on the QoE of ISP customers, revealing the sudden drop of average video download throughput to less than 250~kb/s which hampers even the possibility of watching a  video.

We believe indeed that \tool may serve as a promising tool for ISPs, network administrators, developers and researchers to monitor the YouTube CDN and the traffic it generates. Moreover, thanks to its design, \tool allows to quickly pinpoint the \nodes involved in the changes, thus accelerating the troubleshooting procedures. For instance, ISPs may rely on \tool to quickly react when they observe changes in YouTube CDN which may impair the QoE of their customers, or to design traffic engineering algorithms to automatically adapt the network routing to the changes \tool points out.

While we engineer \tool to target YouTube CDN monitoring, we believe that the \distance notion we introduce in this paper constitutes a more general framework that has the potential to open the usage of unsupervised algorithms for anomaly detection problems in general.

The remainder of this paper is structured as follows: Sec.~\ref{sec:related} discusses the related work. Sec.~\ref{sec:dataset} describes the details of the datasets we use to validate \tool, and shows the dynamicity of YouTube cache selection policies. Sec.~\ref{sec:method} presents our methodology and introduces the \distance. Sec.~\ref{sec:results} presents our results: First, we evaluate the sensitivity of \tool's parameters, and, second, we show how effective \tool is at pinpointing changes in YouTube CDN employing our traces. Finally, Sec.~\ref{sec:conclu} concludes the paper.

%
%
%
%
\section{Related Work}
\label{sec:related}
YouTube had been the subject of study in many papers from different perspectives: from users' behavior~\cite{Gill_2007}
to the social network~\cite{Xu_2008},
from video popularity dynamics~\cite{Figueiredo_2011}
to protocols aspects~\cite{Alcock_2011,Krishnappa_2013}.

Considering the YouTube delivery infrastructure, a large body of work verified its evolution over time~\cite{ CalderIMC13, Adhikari2012_infocom, Torres_2011, Cas2014, Adhikari_2010,  Hossfeld_2013}.
They show a highly dynamic system which keeps changing over time due to continuous upgrades in the infrastructure~\cite{CalderIMC13,Adhikari2012_infocom} or due to the dynamicity of the cache selection policies~\cite{Torres_2011,Cas2014,Adhikari_2010,Hossfeld_2013}.
Some of the findings are already outdated. For instance, the load-balancing policy based on HTTP redirections which is described in~\cite{Torres_2011,Hossfeld_2013} is no longer in place, and YouTube dismissed the naming scheme described in~\cite{Adhikari2012_infocom} at the end of 2011.
In this work, we do not aim to offer an updated view of YouTube. We rather aim at offering a methodology that allows to automatically identify changes in both the infrastructure, e.g., the appearance of new \nodes, and in the day to day management of the infrastructure, e.g., a change in the load-balancing algorithm that may affect millions of customers.


In some sense, our contribution is in line with the body of works focusing on anomaly detection, for which~\cite{PatchaCN2007,Chandola2009} offer good surveys. Most of the works in this area target anomalies in a security context, e.g., the design of intrusion detection systems. Most of them fall in the ``supervised'' category, i.e., given a baseline is built, the proposed methodologies highlight deviations from it.
To the best of our knowledge, only~\cite{ArgusInfocom2012} targets large scale anomaly detection in operational networks. However, it presents a supervised system, which relies on data from passive probes, topology information and routing tables to feed a classic forecasting system, which finally compares its predictions to the actual measurements to pinpoint deviations.
\tool, on the other hand does not assume any knowledge of a baseline, and leverages unsupervised algorithms to automatically unveil	 changes. We specifically design it to target the YouTube CDN, for which the ground truth is a moving target that is very difficult to know.

The application of unsupervised learning techniques to get insights about the network traffic is not new. \cite{mcgregor2004flow} is one of the first works in the context of traffic classification, while \cite{erman2006traffic, wang2010automatic} compare the accuracy of different and standard unsupervised algorithms. \cite{munz2007traffic} proposes a flow-based anomaly detection algorithm based on k-means, while~\cite{TorresSigmetrics2009} uses DBSCAN to identify anomalous clusters. In all the cases, clustering is used to study the same given dataset. To the best of our knowledge, no approaches have been proposed to identify anomalies by comparing the results of clustering applied to different data sources (e.g., different datasets, different time snapshots, etc.). Only~\cite{goldberg2010measuring} aims at measuring similarity between sets of overlapping clusters from complex networks, in which groups of nodes form tightly connected
units that are linked to each other. Since points are not embedded in a metric space, they define ad-hoc distances. \tool operates in a geometric space where we can exploit the concepts of density and centroid of a cluster to simplify the comparison among two different datasets.

\tool differs also from techniques for the tracking of moving clusters and objects as~\cite{kalnis2005discovering, li2010swarm}. Indeed, their goal is to track the movements of the same clustered objects over time, e.g., a group of migrating animals. On the contrary, \tool has no insights about the CDN infrastructure and it cannot track single objects, which may disappear and reappear freely. 

Finally, other approaches as~\cite{KiferVLDB04} aim at measuring the distance among different time snapshots by considering the sample distributions obtained from them. However, directly relying on distributions to perform the comparison considerably complicates the detection of the \nodes behind the changes. Instead, \tool extracts and compares clustering patterns, which are simpler to process in an automatic manner, and allow to immediately pinpoint the \nodes (i.e., the clusters) responsible for possible deviations.
%
%
%
%
\section{Datasets}
\label{sec:dataset}

\begin{figure}[t!]
\centering
    \includegraphics[trim=0cm 3cm 0cm 0cm,width=0.9\columnwidth]{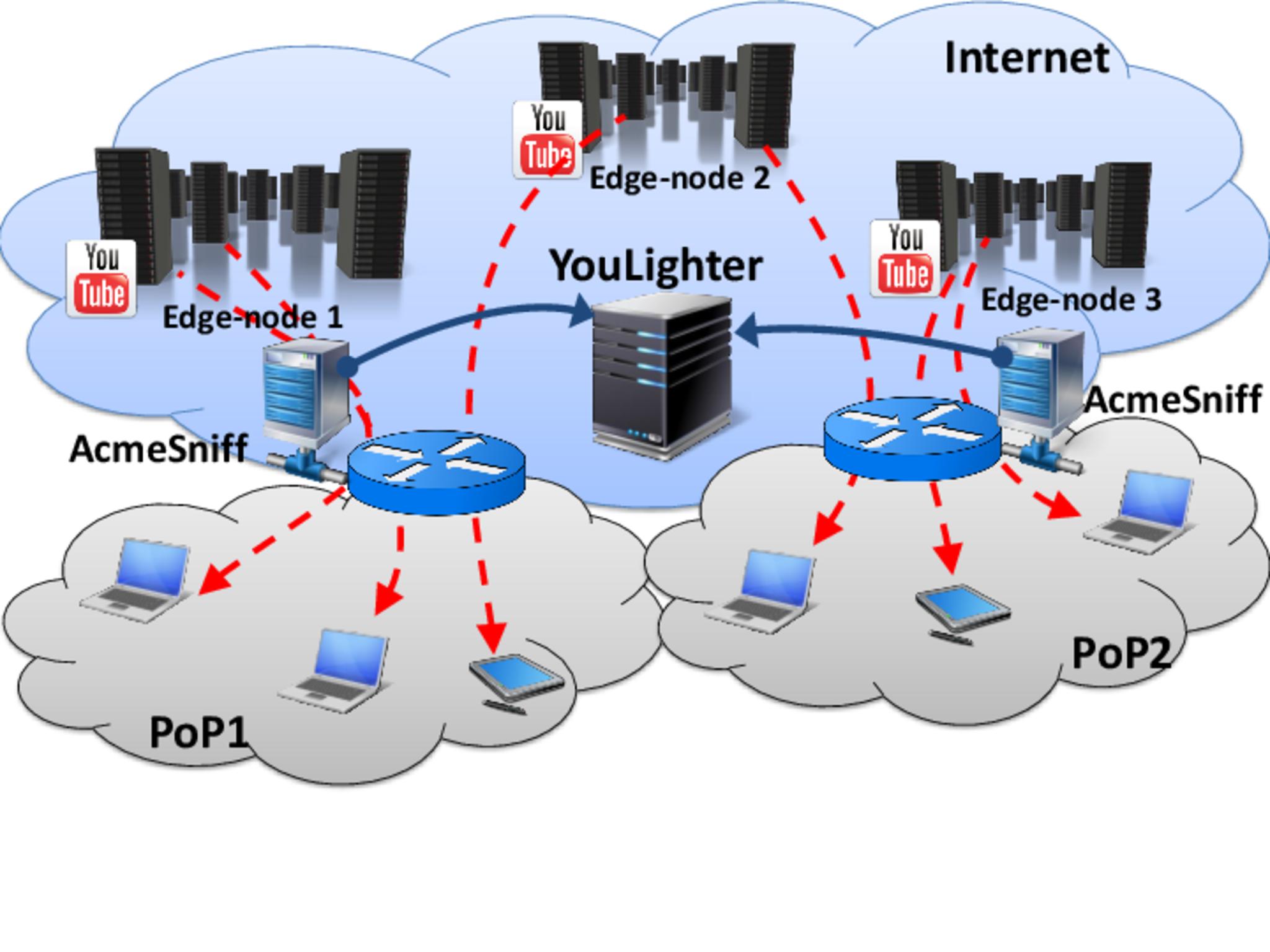}
    \caption{The traffic monitoring setup we employ for this paper.}
    \label{fig:tstat}
\end{figure}

We assume the ISP has instrumented the network with passive probes, which collect statistics from flows carrying YouTube videos.
In this work, we rely on passive probes running \tstat~\footnote{\url{http://AcmeSniff.org/AnonimizedURL}} that we install in Points-of-Presences (PoPs) of operational networks, as depicted in Fig.\ref{fig:tstat}. Clients are located in one PoP, and connect to the backbone via a router, where \tstat monitors the traffic.
\tstat observes packets, rebuilds each TCP flow, tracks it, and at the end of flow, logs detailed statistics.
\tstat can classify TCP flows that carry YouTube videos.
For each request, it logs i) the anonymized client IP address, ii) the server
IP address, iii) the hostname of the server, 
iv) the TCP minimum Round Trip Time (RTT),\footnote{The RTT is measured as the time between
the client data segment and the corresponding server acknowledgment observed at the vantage point.  For each TCP connection, minimum RTT is computed among all valid samples.} v) the IP Time-To-Live (TTL) of packets received by the client in the PoP, vi) the amount of bytes the clients send and receive, vii) the average download throughput, and viii) the time at which the TCP connection starts. Note that \tstat can compute all these metrics considering only TCP segments, and do not require access to application payload. This avoids any privacy issues.

\begin{table}[h]
\centering
\resizebox{1.0\columnwidth}{!}{\centering
\begin{tabular}{c|c|c|c|c}
\textbf{Trace}      & \textbf{Period}       & \textbf{Volume}    & \textbf{Videos} & \textbf{Caches}      \cr\hline
\TApub\     & 01/04/2013 - 28/02/2014           & 138.7~TB      &  2,892,452        & 8,664 \cr
\TApri\     & 01/04/2013 - 28/02/2014           & 152.9~TB      & 2,848,625         & 8,899 \cr
\TBpub\     & 01/04/2013 - 28/02/2014       & 134.8~TB      & 2,711,179         & 9,028 \cr
\TC\            & 01/03/2014 - 17/07/2014           &  48.3~TB  &   305,802         & 3,755  \cr
\end{tabular}
}
\caption{Traces considered in this study.}
 
\label{tab:desc-traces}
\end{table}

We have been collecting traffic logs since April 2013 by monitoring the traffic users generate when accessing the Internet. We instrument four different PoPs. Three of them are located in PoPs of the same ISP and in two different cities of the same country. We install the fourth one in a PoP of a different ISP in a second country.
Tab.~\ref{tab:desc-traces} describes, for each trace (or dataset), the time period, the total downloaded volume, the number of unique videos and the number of YouTube servers we observe. Notice that in total we monitor the activity of more than 32,000 customers, and the maximum number of caches that ISP1 customers used at least once is $\sim$9,000.


\subsection{YouTube Cache Naming Structure}
\label{sec:prel}

We find that the YouTube infrastructure described in~\cite{Adhikari2012_infocom} is no longer in use.
Since 2012, YouTube server hostnames are in the form {\texttt{rx---ABCxxtxx.c.youtube.com}}, where \texttt{x} are numbers, while \texttt{ABC} is a three-letter code reporting the IATA code of the closest airport. For instance
{\texttt{r7---fra07t16.c.youtube.com}} identifies a single cache, in Frankfurt. The hostname resolves to a single IP address, 74.125.218.182 in the example. Thus, we can uniquely identify a cache by its hostname.\footnote{Starting from January 2013, YouTube obfuscates the IATA code using a simple substitution cipher that we were able to break. For instance, \texttt{r7---fra07t16.c.youtube.com} becomes \texttt{r7---sn-4g57kued.c.youtube.com}. From October 2013, the \texttt{youtube.com} domain has been replaced by the \texttt{googlevideo.com} domain. This information can be used to identify YouTube flows even in presence of \cite{BermudezIMC12}.}
All caches co-located in the same \node share the same IATA code. This allows us to get coarse ground truth about the location of servers. However, as we will see, several \nodes can be located in apparently different areas, but share the same IATA code.

We run some active experiments to cross-check if YouTube specializes caches to serve some particular content, and we verify that every cache can serve any video, at any resolution, in any format, e.g., MPG4 or Flash, to any device, e.g., PC, smartphones or tablets.


\subsection{Characterization of the Load Balancing Policies}

\begin{figure}[t!]
\centering
    \includegraphics[width=1\columnwidth]{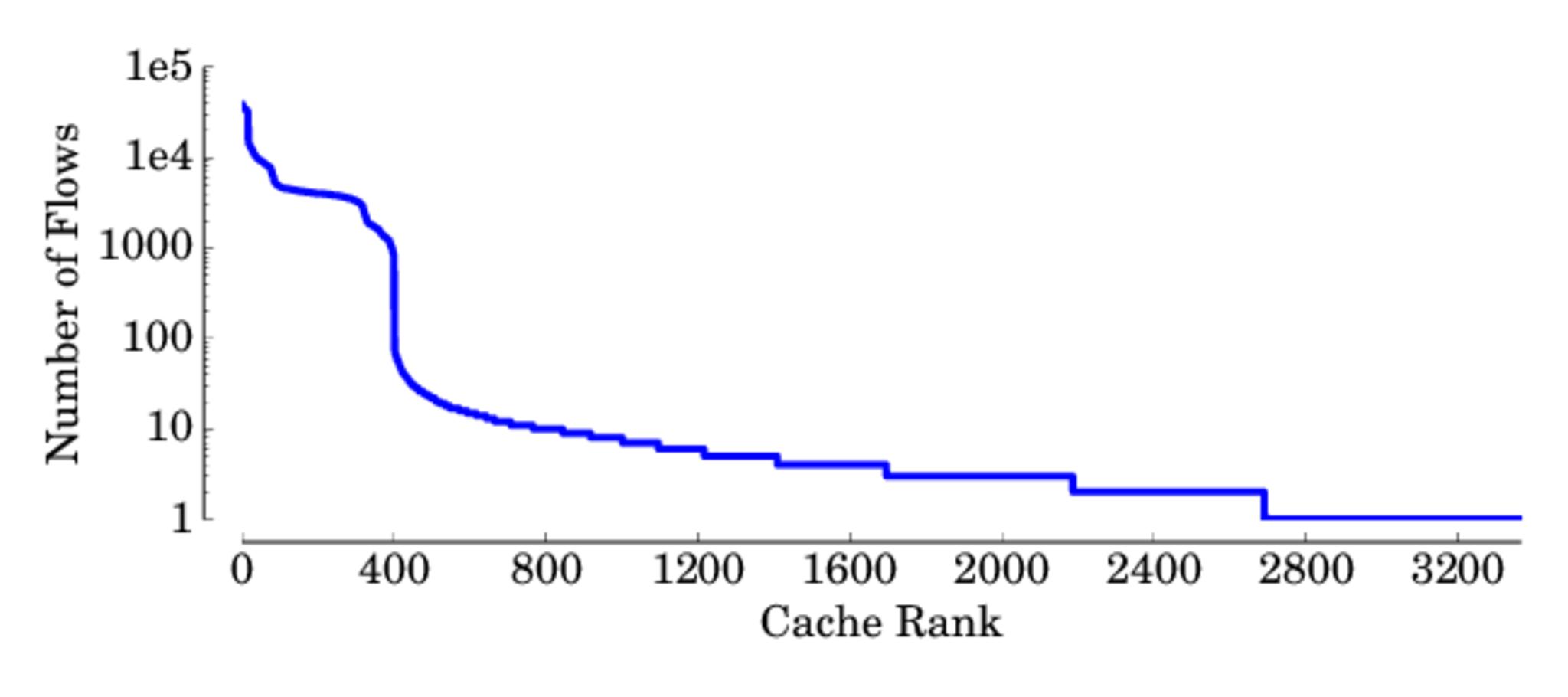}
    \caption{Rank of YouTube caches based on the number of flows. February 2014, \TApub.}
    \label{fig:ip_rank}
\end{figure}

Every time a user starts a video playback, the player starts a progressive download of the video content from the specific cache the system provides in the HTML page.\footnote{Load balancing policies are implemented at application layer. Indeed the web server chooses and encodes the cache hostname directly in the HTML page served to the client.}
We are interested in seeing which are the policies governing the server allocation, such as (i) is there any ``preferred'' group of caches? or (ii) are those stable over time?
Fig.~\ref{fig:ip_rank} reports the rank of YouTube caches based on the number of flows they handle. We consider February 2014 from the \TApub dataset.
First, notice that we observe more than 3,200 caches during one month. Second, the load each cache handles is very heterogeneous; few servers handle lots of requests, but there is a not negligible number of caches that serves a significant portion of flows. For instance, more than 400 caches serve more than 100 videos, and in order to to observe 95\% of requests, one should monitor about 330 caches.

\begin{figure}
\centering
    \includegraphics[width=1\columnwidth]{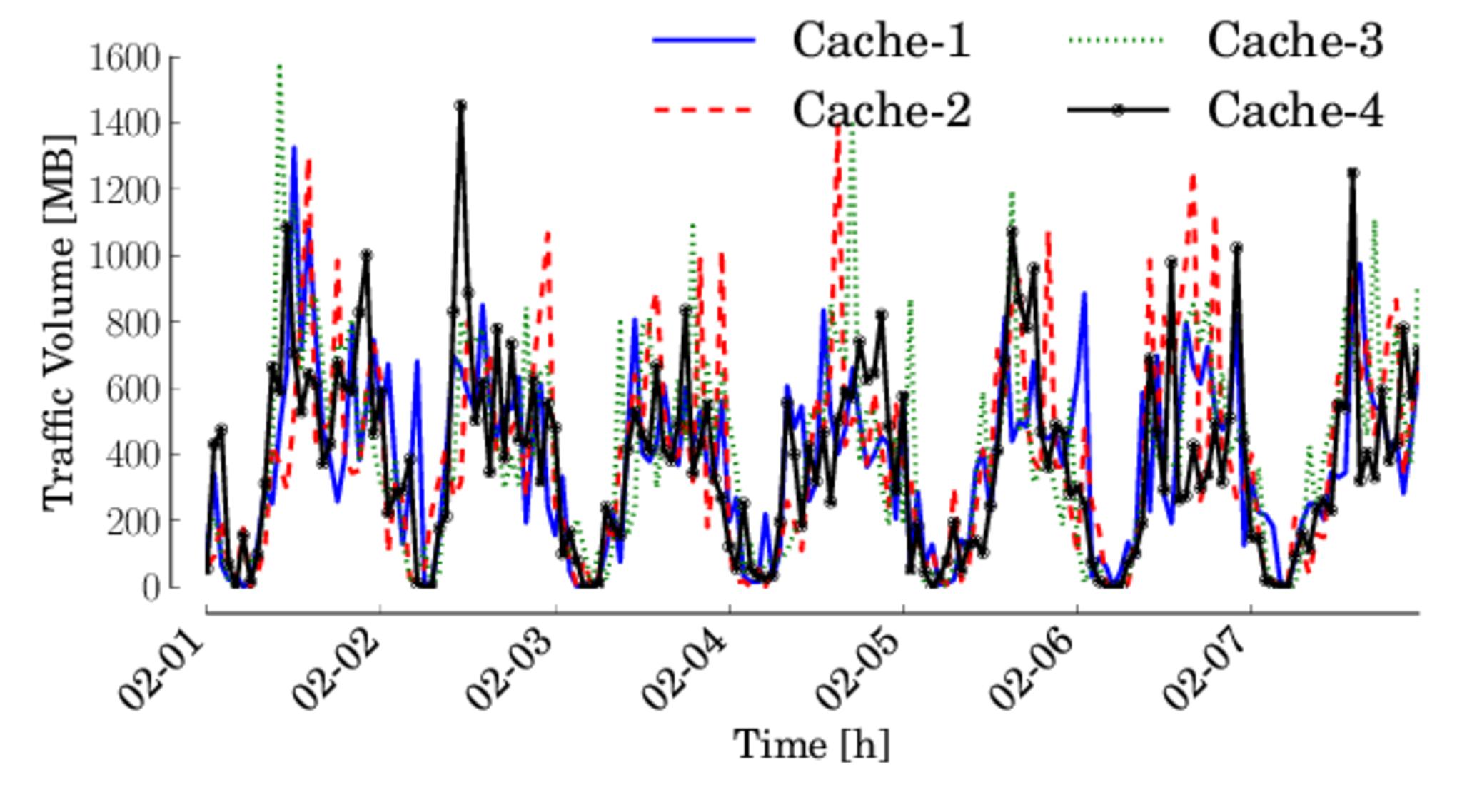}
    \caption{Evolution of the volume of traffic for the four most active caches we observe on February 1st 2014. First week of February 2014, dataset \TApub.}
    \label{fig:tren4}
\end{figure}

We also notice that the rank is extremely dynamic over time. For instance, we pick the four most active caches during the 1st of February 2014 and we report in Fig.~\ref{fig:tren4} the amount of traffic they generate over time for the following seven days. As shown, the amount of traffic a single cache handles changes widely over time, and none of the monitored caches keeps a constant leading position for a long period of time.

\begin{figure}[t!]
\centering
    \includegraphics[width=1.0\columnwidth]{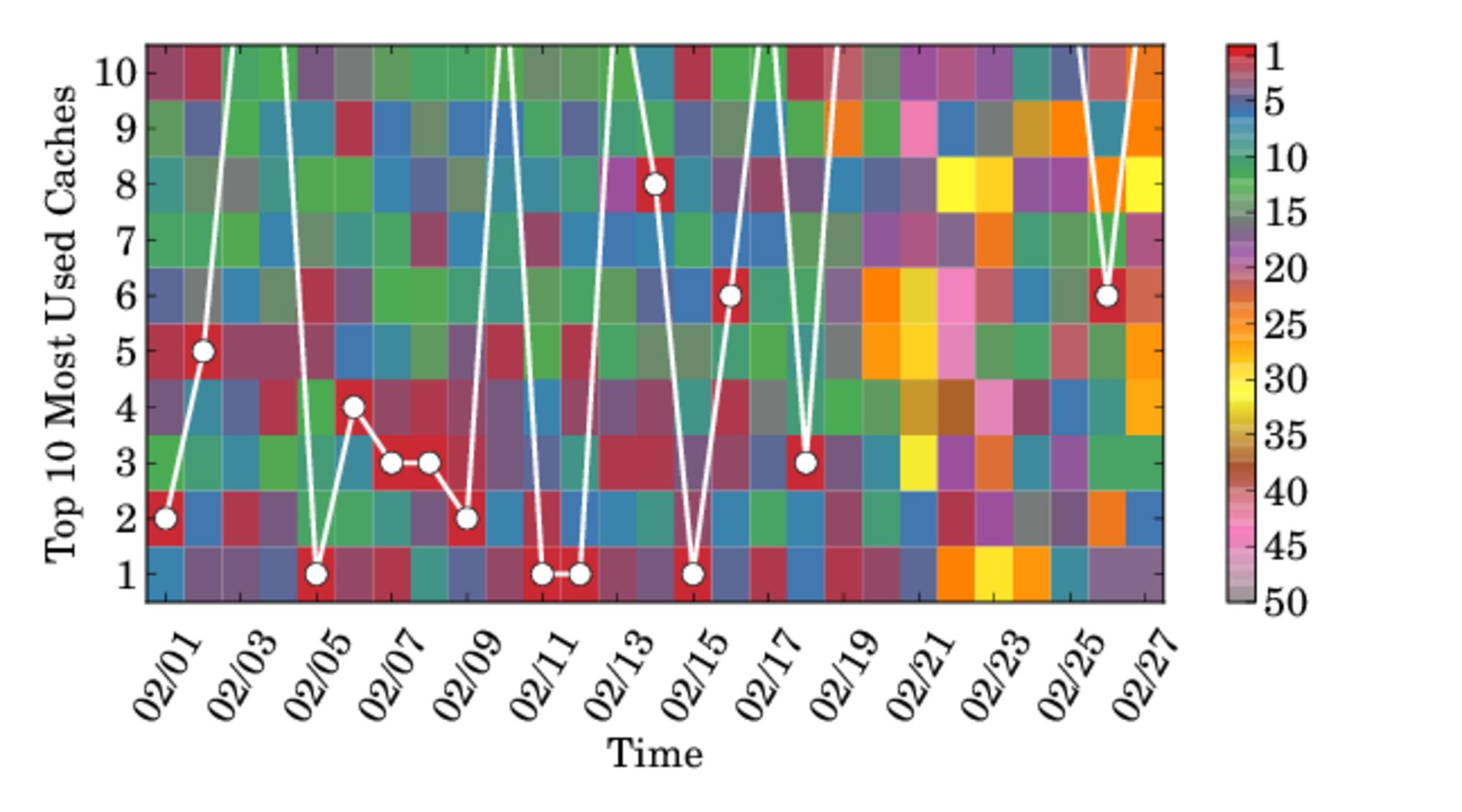}
    \caption{Evolution over time of the rank of the top 10 mostly used caches during February 2014, \TApub. The white dot corresponds to the top cache of each day.}
    \label{fig:ip_monitoring}
\end{figure}

As one may expect this dynamicity to disappear when reducing the focus, we monitor a larger pool of caches as those in the rank in Fig.~\ref{fig:ip_rank}, and we recompute the same rank on a daily basis. Then, we represent it using different colors in Fig.~\ref{fig:ip_monitoring}. Each row represents the rank of the same cache for different days in February.
In case the rank is stable, one would expect a row (a cache) to always assume the same color (rank). Fig.~\ref{fig:ip_monitoring} shows exactly the opposite. Indeed the top daily cache (red square, highlighted by the white dot) randomly changes every day (white line). Sometimes, the most used cache in a day is not among the top-10 cache of the month (line jumps outside). The top-10 caches in the monthly rank drops below the 50th place during some days (gray color). Similarly, in the first 19 days of February, the top-10 caches are concentrated in the first 20 rankings; However, starting from February 20th they fall around the 30th position (notice the concentration of yellow and orange).

This shows that the server allocation policies adopted by YouTube spread the load over several hundreds of caches, and the choices are extremely dynamic over time if we observe with the fine grained granularity of a single cache.

Since caches inside the same \node are all equivalent, the intuition is to observe the system using the coarse granularity offered by \nodes.
However, \nodes are unknown, they can change over time due to system upgrade or redesign, and information that could be available (e.g., the IATA code) may be not reliable, or may be removed by YouTube.
In the following, we design an unsupervised clustering algorithm to automatically identify \nodes from just the observation of traffic flows.

%
%
%
%
\section{Methodology}
\label{sec:method}

\begin{figure}[t!]
\centering
     \vspace*{-0.1cm}
    \includegraphics[width=1\columnwidth]{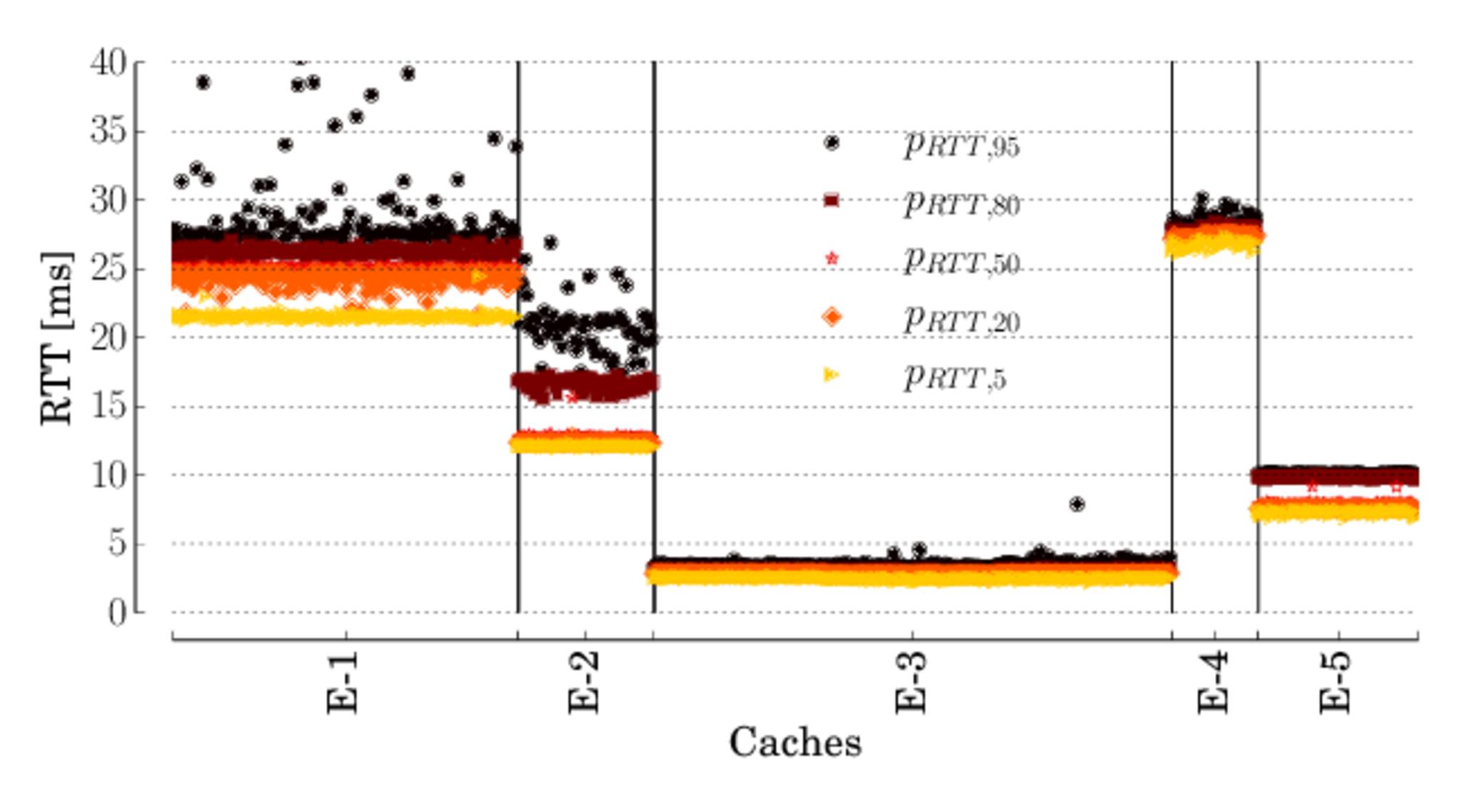}
     \vspace*{-0.5cm}
    \caption{Example of per cache RTT percentiles. Caches sorted by IP address, and grouped by (anonymized) IATA code.}
    \label{fig:perc1}
     \vspace*{-0.3cm}
\end{figure}

Intuitively, the path between two caches in the same \node and clients in the same PoP
exhibits the same properties, e.g., same RTT. Conversely, the path between two caches in different \nodes should present different RTT. This intuition is corroborated in Fig.~\ref{fig:perc1} which depicts the 5th, 20th, 50th, 80th, and 95th percentiles of the per-cache RTT distribution. We identify caches with their IP address, and then we order and group them into \nodes using the IATA code as ground truth so that caches belonging to the same \node appear one close the other. Five \node are present, E-1 to E5. Each hosts a variable number of caches, with E-3 being the largest.
As shown, the caches in the same \node exhibits very similar RTT percentiles, suggesting that we can identify clusters of caches by considering the RTT as a feature.

\subsection{Multi-dimensional Clustering}
\label{sec:clust}

We leverage above intuition to design a clustering algorithm to automatically find homogeneous groups of caches.
We use some ingenuity to characterize the path from client to each cache, and then to cluster caches that exhibits similar paths.
We can split the process of our methodology into the following steps:

\noindent{\bf Step 1 - Passive monitoring of YouTube video flows}:
As described in Sec.~\ref{sec:dataset}, a passive probe provides the continuous collection of YouTube traffic logs. We log each the metadata of each TCP connection, and we store logs in a database for further processing.
\\
\noindent{\bf Step 2 - Measurement consolidation and filtering}:
To ease the monitoring procedure, we use a batch processing approach that considers time windows of size $\Delta T$. Thus, every $\Delta T$ we generate a ``snapshot'', and we aggregate and process measurements in it. In the following, we indicate the $n$-th snapshot as a superscript when needed, e.g., $a^{(n)}$ indicates the metric $a$ at snapshot $n$.

We identify each cache $x$ by its IP address. We then group all flows in the same snapshot with the same server IP address to obtain a table where columns correspond to the metric (e.g., RTT, TTL, transmitted packets, etc.),
and each row corresponds to a sample, i.e., the tuple of measured values observed within a TCP flow.

Since we are interested in the active caches, we discard those with less than $MinFlow=50$ samples.
We define the whole measurement snapshot $n$ as $X^{(n)}$.\\
\noindent{\bf Step 3 - Feature selection and data normalization}:
Next, we apply a feature selection driven by domain knowledge to select the set $\mathcal{M}$ of {\it metrics}. In particular, as we are interested in grouping caches according to the path properties, we choose $\mathcal{M}=\{RTT, TTL\}$. Then, for each cache $x$ in the snapshot $X$, and for each metric $m\in \mathcal{M}$, we generate an empirical distribution.
From the distribution, we extract the vector $P_m(x)=\left(p_{m,1}(x),p_{m,2}(x),\ldots,p_{m,k}(x)\right)$ containing $k$ percentiles of $m$ for cache $x$.
We thus standardize percentiles following a simple normalization:
\begin{eqnarray}
	min_m = \min\left(p_{m,i}(x) \ \forall x \in X,  \forall i=1, \ldots, k\right) \\
	max_m = \max\left(p_{m,i}(x) \ \forall x \in X,  \forall i=1, \ldots, k\right) \\
	\bar{p}_{m,i}(x) = \frac{p_{m,i}(x) - min_m}{max_m - min_m} \label{eq:norm}
\end{eqnarray}
Intuitively, Eq.(\ref{eq:norm}) normalizes the percentiles of metric $m$ so that $\bar{p}_{m,i} \in [0,1]$.

At last, $\bar{P}_m(x)=\left(\bar{p}_{m,1}(x),\bar{p}_{m,2}(x),\ldots,\bar{p}_{m,k}(x)\right)$ represents the standardized vector of {\it features} for the metric $m$ for server $x$.
Recalling that $\mathcal{M}=\{RTT,TTL\}$, we identify each cache $x\in X$ with a $2k$-dimensional space of edge 1 by features:
\begin{equation}
 \bar{x} = ( \bar{P}_{RTT}(x), \bar{P}_{TTL}(x))
\end{equation}
and we transform the original set of caches $X$ into a set of points $\bar{X}=\{\bar{x}\}$.
\\
\noindent{\bf Step 4 - Clustering}:
We employ the density-based DBSCAN algorithm~\cite{ester1996density} to group together servers based on their multi-dimensional features. We choose DBSCAN because (i) it is able to handle clusters of arbitrary shapes and sizes; (ii) it is relatively resistant to noise and outliers; and (iii) it does not require the specification of the number of desired clusters. DBSCAN requires two parameters: $\epsilon$ and $MinPts$. $\epsilon$ determines the maximum allowed distance between any given point in a cluster and its closest neighbor belonging to the same cluster, and $MinPts$ the minimum number of points required to form a cluster. Based on that, it classifies all points as being (i) core points, i.e., in the interior of a dense region; (ii) border points, i.e., on the edge of a dense region; or (iii) noise points, i.e., in a sparsely occupied region. Noise points do not form any cluster, while the algorithm puts in the same cluster any two core points that are within $\epsilon$ of each other. Similarly, any border point that is close enough to a core point is put in the same cluster as the core point.
%
%
The result of this process is a collection $\mathcal{C}$ of clusters $C_j\in\mathcal{C}$, also named as \textit{clustering}:
\begin{equation}
 \mathcal{C} = \{C_j\}=\text{DBSCAN}(\bar{X})
\end{equation}

\subsection{Highlighting Changes with the Constellation Distance}
\label{sec:distance}
We are now interested in tracking the evolution of clusters over time, for which, as we discuss in Sec.~\ref{sec:related}, no known solution is present in the literature. Indeed, it is not obvious how to compare two clusterings $\mathcal{C}1$ and $\mathcal{C}2$ obtained considering two {\it different} datasets, i.e., snapshots in our case.
For instance, i) points that were present in $\mathcal{C}1$ may not be present in $\mathcal{C}2$, and vice versa; ii) points clustered into the same cluster in $\mathcal{C}1$ can now belong to two or more clusters in $\mathcal{C}2$; and iii) the same points that form a cluster in $\mathcal{C}1$ can still form the same cluster, but can be placed in another region in the clustering space in $\mathcal{C}2$. In our case, this corresponds to i) popular caches at snapshot $n$ that are not anymore used at snapshot $n+1$, or ii) some caches at snapshot $n$ that were part of the noise are instead clustered at snapshot $n+1$, or iii) the path to caches suddenly changes at snapshot $n+1$, altering RTT and TTL.

To evaluate the difference among the clustering, we propose a novel methodology that is based on the notion of \distance.

\subsubsection{Constellation}
We first map each cluster into a single {\it star} that summarizes it. Given a cluster $C\in\mathcal{C}$, we consider the centroid, or geometric center,
$\hat{x}$ whose components $\hat{p}_{m,i}$ in the $i$ percentile of feature $m$ are:
\begin{eqnarray}
 \hat{p}_{m,i} = \frac{1}{|C|} \sum_{x\in C}{renorm({p}_{m,i}(x))}
\end{eqnarray}
All stars then form a {\it constellation} $\hat{\mathcal{C}}=\{\hat{x}\}$.
The $renorm()$ function eventually considers the re-normalization of features that can be needed if points in $\mathcal{C}1$ and $\mathcal{C}2$ went through different standardization processes.
In our case, assuming $\mathcal{C}1=\mathcal{C}^{(n)},\ \mathcal{C}2=\mathcal{C}^{(n+1)}$, from Eq.(\ref{eq:norm}) for each $m\in \mathcal{M}$ we have:
\begin{eqnarray}
	Min_m = \min\left(min_m^{(n)},min_m^{(n+1)}\right) \\
	Max_m = \max\left(max_m^{(n)},max_m^{(n+1)}\right) \\
	renorm_m(a) = \frac{a - Min_m}{Max_m - Min_m}
\label{eq:renorm}
\end{eqnarray}

\subsubsection{Astral Distance}
Given a star $\hat{x}$ and a constellation $\hat{\mathcal{C}}$, we define the {\it Astral Distance} ($AD$) 
as the distance between $\hat{x}$ and its closest star in $\hat{\mathcal{C}}$.
Specifically, we compute the closest star $\hat{y}^* \in \hat{\mathcal{C}}$  such that $d(\hat{x},\hat{y}^*) \leq d(\hat{x},\hat{y}) \,\, \forall \hat{y} \, \in \, \hat{\mathcal{C}}$. $d(x,y)$ can be any distance metric that is valid in the feature space. In this work, we use the classic Euclidean distance.
Thus, the Astral Distance $AD$ of the star $\hat{x}$ from stars in $\hat{\mathcal{C}}$ is
\begin{eqnarray}
 AD(\hat{x}, \hat{\mathcal{C}}) = \min_{\hat{y} \in \hat{\mathcal{C}}} d(\hat{x}, \hat{y})
\end{eqnarray}

Hence, the Astral Distance couples stars according to a nearest neighbor principle. 

\subsubsection{Constellation Distance} At last, we define the \distance\ - $CD$ - as the sum of the Astral Distances among every star in the clusterings. Since the number of clusters in $\hat{\mathcal{C}}1$ and $\hat{\mathcal{C}}2$ may be different, we need to symmetrize the definition:
\begin{equation}
\begin{split}
CD(\hat{\mathcal{C}}1,\hat{\mathcal{C}}2) =
     \sum_{\hat{x}\in \hat{\mathcal{C}1}} AD(\hat{x}, \hat{\mathcal{C}}2)\  + \ 
             \sum_{\hat{x}\in \hat{\mathcal{C}}2} AD(\hat{x}, \hat{\mathcal{C}}1)
\end{split}
\label{eq:constDist}
\end{equation}

\begin{figure}[t!]
\centering
    \includegraphics[trim=0cm 8.5cm 0cm 0cm, width=1\columnwidth]{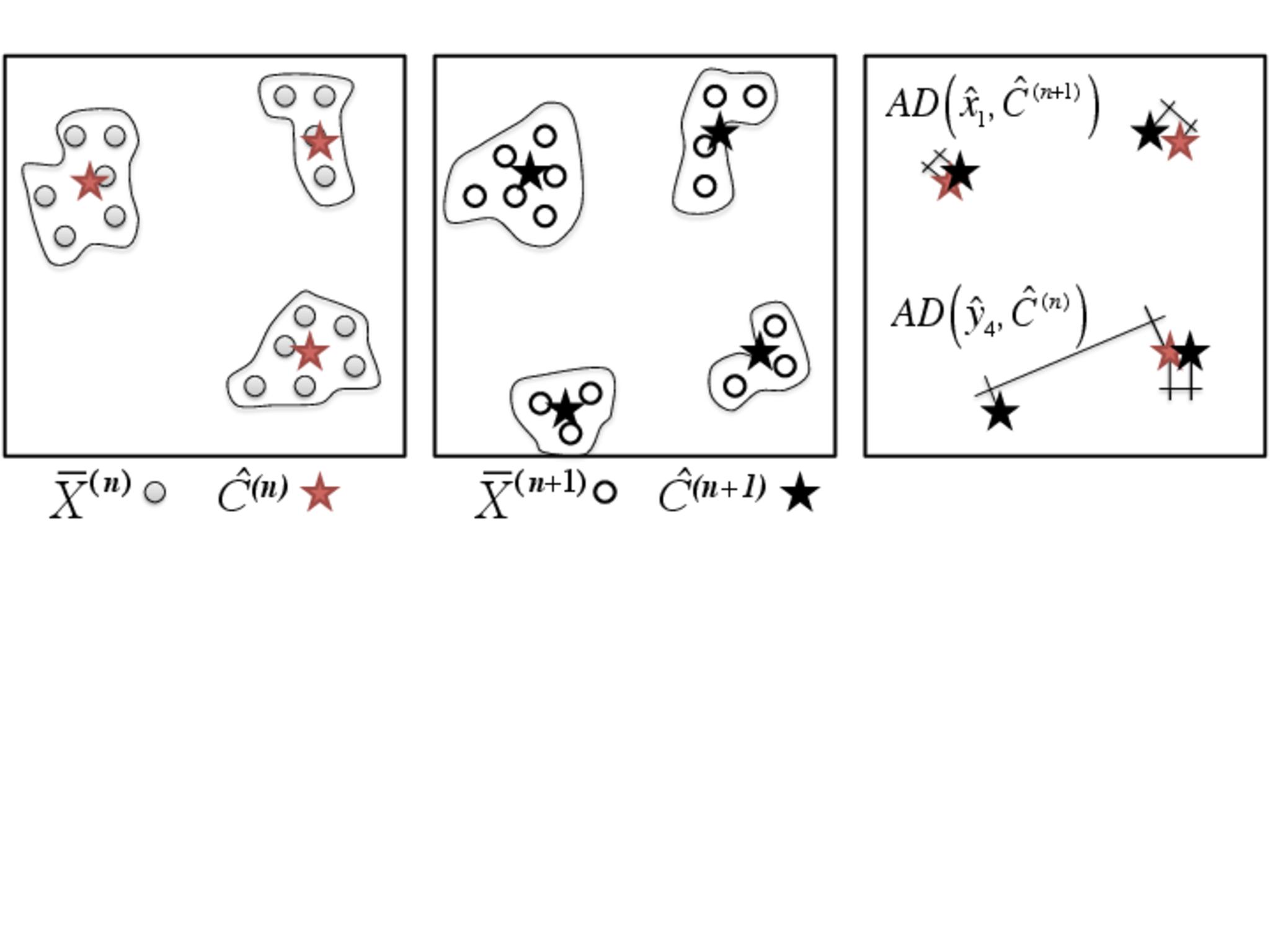}
    \caption{Example of Clusterings, Constellations and Astral Distance computations.}
    \label{fig:distance}
\end{figure}

Fig.~\ref{fig:distance} depicts the \distance computation considering a 2-dimensional space. From left to right, DBSCAN first clusters the points (grey dots for the first snapshot, white for the second). Then, stars emerge to form the constellations, and we compute the Astral Distance for each star. Finally, the \distance is the sum of all Astral Distances.

In the following, we consider two subsequent snapshots $n$, and $n+1$, compute the clustering $\mathcal{C}^{(n)}$ and $\mathcal{C}^{(n+1)}$, then extract the constellations  $\hat{\mathcal{C}}^{(n)}$ and $\hat{\mathcal{C}}^{(n+1)}$, and finally compute their distance $CD\left(\hat{\mathcal{C}}^{(n)},\hat{\mathcal{C}}^{(n+1)}\right)$.

As we discuss in Sec.~\ref{sec:related}, to the best of our knowledge we are the first to propose an approach to quantify the similarity among different clustering results. We note that we can base the \distance on other similarity metrics different from the Euclidean distance, e.g., the well known Cosine Similarity. 
However, as we show in Sec.~\ref{sec:observation} using the Euclidean distance lets the \distance to inherit linear properties, and therefore to vary proportionally with size of the changes. Observe also that the design of the \distance offers a nice property that is particularly desirable for troubleshooting purposes. In particular, the \distance, which is a simple sum of Euclidean distances, lets us immediately pinpoint the stars responsible for changes in the constellation. As we show in Sec.~\ref{sec:highlight}, this aspect is crucial, as it allows us to design an automatic procedure that i) captures changes in YouTube CDN infrastructure, and ii) highlights the \nodes involved in these changes.

\begin{figure}[t!]
\centering
\hspace*{-0.2cm}
\subfigure[{$|\hat{\mathcal{C}}1| = |\hat{\mathcal{C}}2|$}]{
	\label{fig:equal}
	\includegraphics[width=0.47\columnwidth]{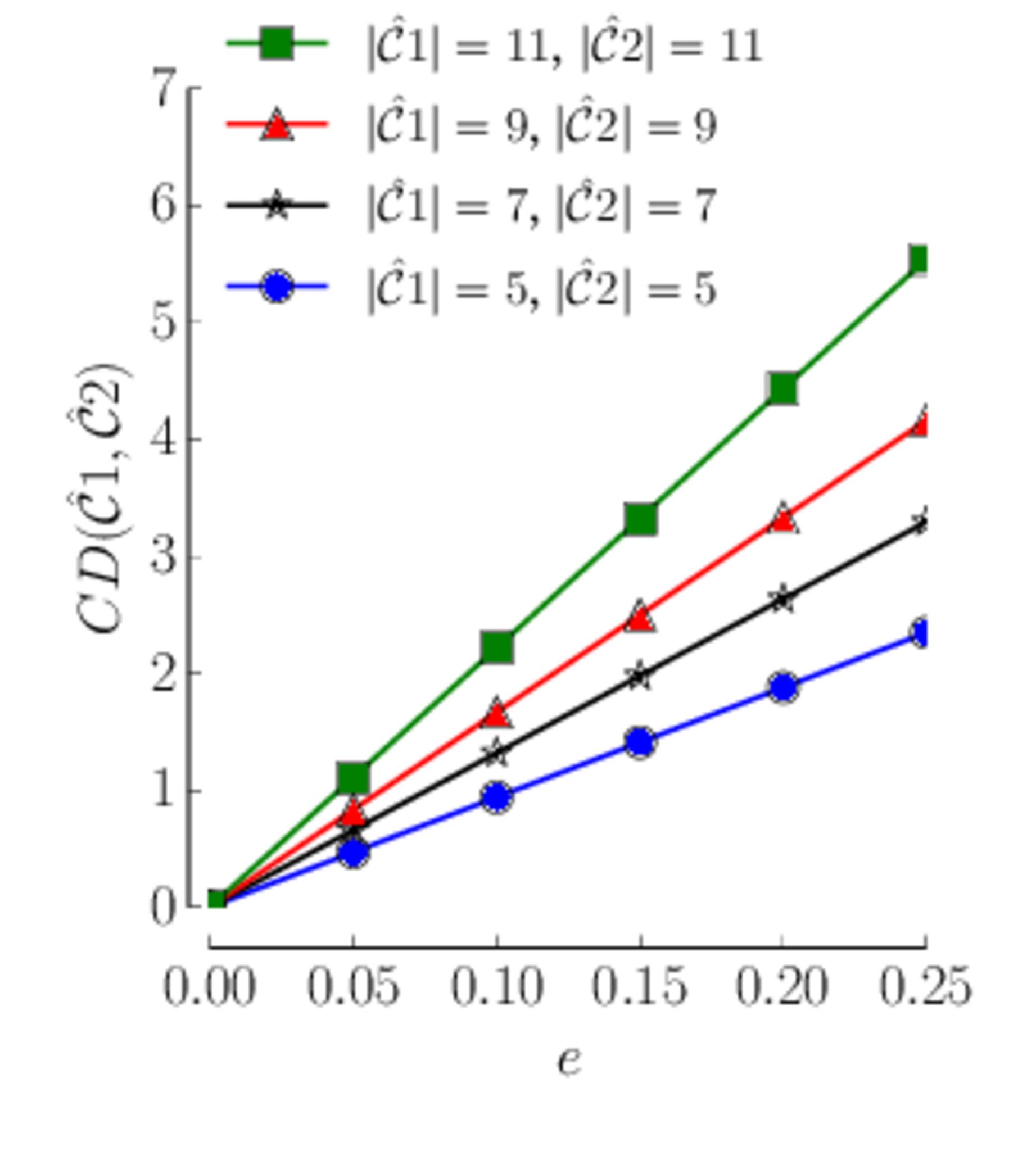}
}
\subfigure[{$|\hat{\mathcal{C}}1| < |\hat{\mathcal{C}}2|$}]{
	\label{fig:extra}
	\includegraphics[width=0.47\columnwidth]{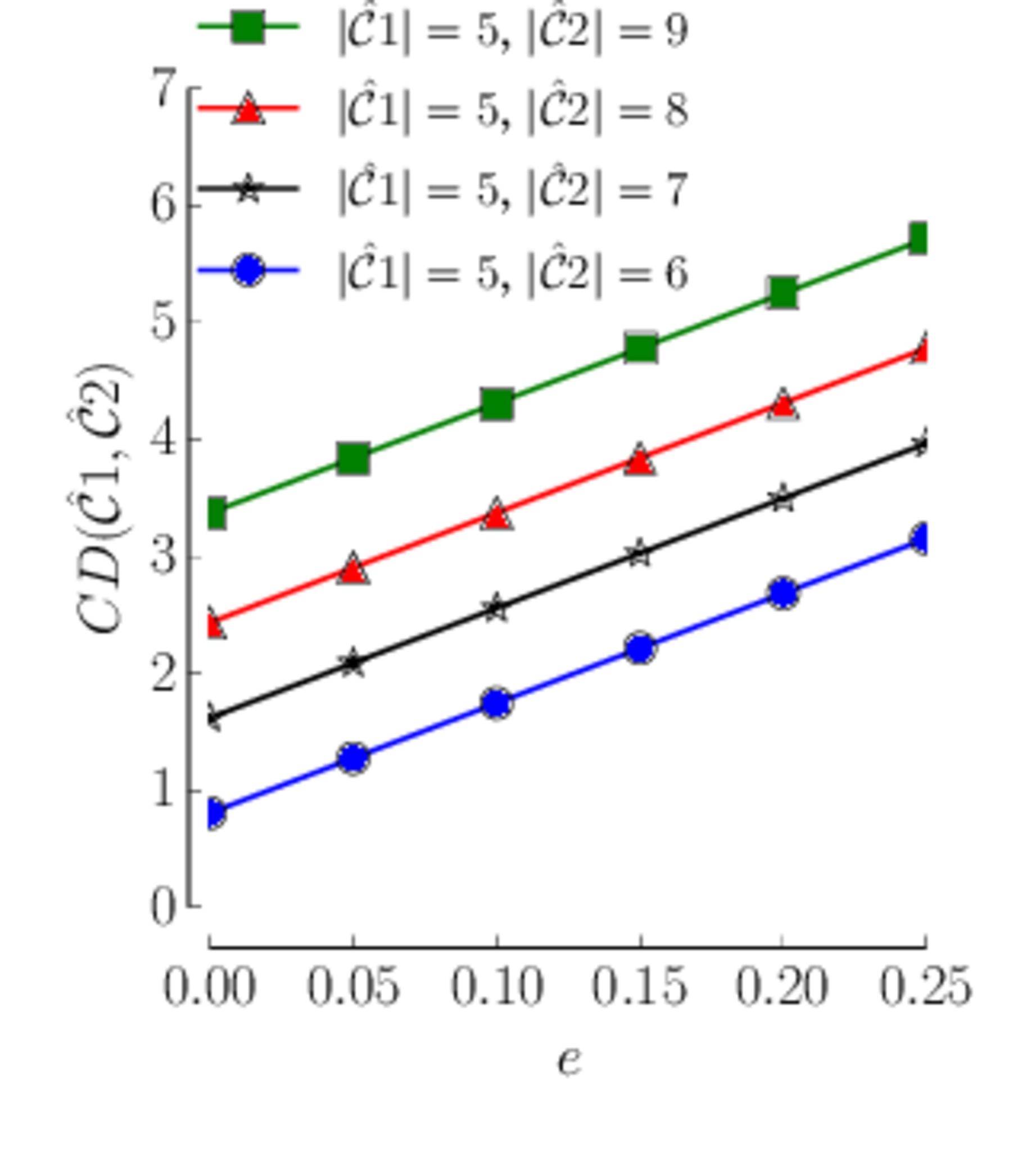}
}
\caption{\distance for increasing noise $e$, and constant or increasing number of stars.}\label{fig:CD}
\end{figure}

\subsection{Observations about the \distintitle}
\label{sec:observation}

We run some numerical evaluation to gauge how the \distance changes with respect to changes in the input data. We consider two main sources of changes: i) stars that simply move from their position, and ii) the birth of new star reflecting the generation of a new cluster in the data.

For the first scenario, we generate a random constellation $\hat{\mathcal{C}}1$ of $N=|\hat{\mathcal{C}}1|$ stars. We randomly place stars in the unitary hypercube of edge 1 in $\mathbb{R}^{N}$ according to a uniform distribution. Then, we generate constellation $\hat{\mathcal{C}}2$ by taking the centroids in $\hat{\mathcal{C}}1$, and repositioning them in a random sphere of radius $e$ centered in the centroid original position. Finally, we compute $CD(\hat{\mathcal{C}}1,\hat{\mathcal{C}}2)$.
We repeat the experiment for $100$ times, and average the obtained values.
Fig.~\ref{fig:equal} reports the average \distance for increasing values of $e$, and for different values of $N$. As expected, curves pass through the origin, and linearly grow with $e$. The larger is $N$, the higher is the average \distance.

For the second case, we run the same experiment while also increasing the number of stars. Thus
$|\hat{\mathcal{C}}1| < |\hat{\mathcal{C}}2|$.
Fig.~\ref{fig:extra} shows the results. Notice the nice property of the \distance for which the birth of new stars causes the \distance to grow by a factor that is proportional to the number of new stars. This is due to definition in Eq.(\ref{eq:constDist}) in which no normalization is present.
This property is important, as it lets the \distance nicely highlight the sudden birth (or death) of stars.



%
%
%
%
\section{Results}
\label{sec:results}

In this section we first assess and tune the performance of DBSCAN in order to identify \nodes.
We next run \tool over a longitudinal dataset to show its ability to highlight sudden changes in the YouTube CDN.

\subsection{DBSCAN performance}
\subsubsection{Clustering Performance Metrics}
We first evaluate the impact of the parameter settings on the DBSCAN clustering results. In particular, we aim to understand how good is the matching between the clustering DBSCAN returns and the \nodes we observe in the measurements. To perform this analysis, we consider the snapshot $X$ from November 4th to November 10th, 2013, in trace \TApub. We manually inspect the dataset, and, guided by the IATA codes, we assign each cache a label corresponding to the \node in the YouTube CDN. We manually cross-check labels by inspecting server IP addresses and subnets, RTT and TTL distributions to verify the accuracy of the labels. The result is a ground truth label, GT-label, that we assign to each cache.
In total we find $|X|=620$ caches serving more than $MinFlow=50$ flows, and belonging to 6 \nodes, each identified by a different GT-label. Hence, the number of GT-labels is $N_{GT}=6$.

We then run DBSCAN as described in Sec.~\ref{sec:clust}, obtaining the clustering $\mathcal{C}$.
Let $N_C=|\mathcal{C}|$ be the number of clusters. We next use the GT-labels to assign a label
to caches by using a majority-voting scheme: For each cluster ${C}_j \in {\mathcal{C}}$, we assign all caches $x\in {C}_j$ the most frequent GT-label observed in ${C}_j$.
Caches whose assigned label matches the GT-label are the so called True Positives (TP), whose number is $N_{TP}$. Conversely, caches whose assigned label is different from their GT-label are False Positives (FP), whose number is $N_{FP}$. $|X|=N_{TP}+N_{FP}$.
We compute the set of distinct labels assigned to clusters in ${\mathcal{C}}$, whose number is $N_L\leq N_{GT}$.
We do not assign any label to the caches which DBSCAN classifies as noise points.

\begin{figure}[t!]
\begin{center}
 \includegraphics[trim=0cm 3.5cm 0cm 1cm, width=1.0\columnwidth]{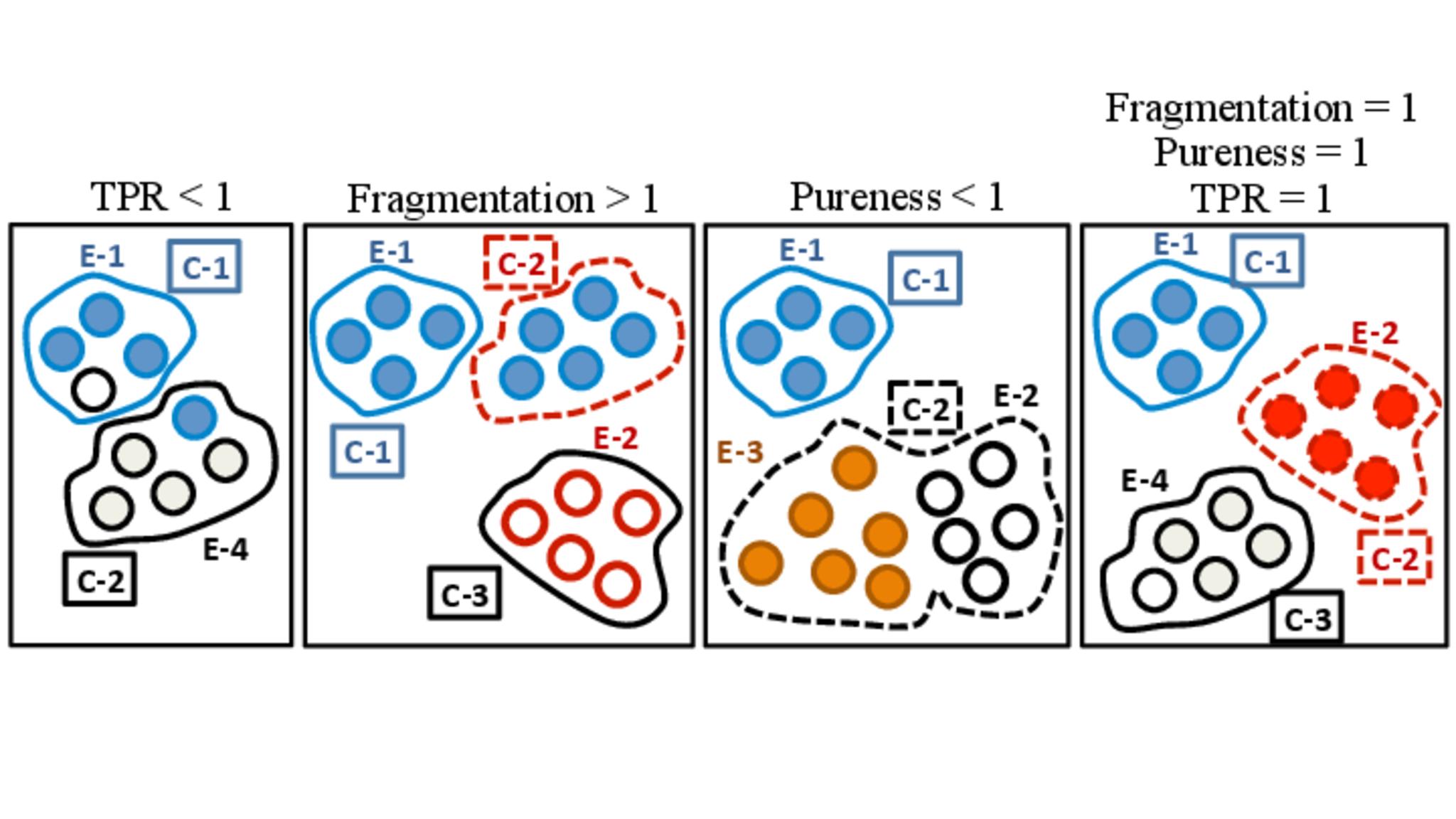}
 \caption{Examples of patterns for which the \textit{True Positive Rate} , the \textit{Fragmentation Index}, and the \textit{Pureness Index} are not equal to 1, and the optimal case in which they are all equal to 1. Color represent the GT-label.}
  \label{fig:indexes}
\end{center}
\end{figure}

To validate the clustering we obtain with DBSCAN, we compute the followings indices:
\begin{equation}
 \mbox{TPR} = \frac{N_{TP}}{|X|}, \,\,
 \mu = \frac{N_C}{N_{L}}, \,\,
 \phi = \frac{N_L}{N_{GT}}
\end{equation}\label{eq:tp_measure}
\begin{enumerate}
\item The True Positive Rate ($TPR\leq 1$) is the ratio between TP and the number of samples in the experiment.
$TPR=1$ means that all labels are identical to the GT-label. $TPR<1$ indicates the presence of i) mislabelled caches (or FP), or ii) noise points (unlabeled points). Leftmost sub-figure in Fig.~\ref{fig:indexes} reports a simple example where the clustering algorithm mislabels a cache for both the GT-labels E-1 and E-2, thus leading to $TPR < 1$. Colors represent the GT-label.

\item The \textit{Fragmentation Index} ($\mu\geq 1$) captures the case when more clusters share the same GT-label.
When $\mu=1$, the number of clusters is identical to the number of GT-labels and DBSCAN assigns each cluster a different GT-label. When $\mu>1$ instead, we have more clusters which share the same GT-label, i.e., DBSCAN splits an \node into two or more clusters. Second sub-figure in Fig.~\ref{fig:indexes} reports an example where the clustering algorithm splits \node E-1 in two different clusters, C-1 and C-2, thus leading to $\mu > 1$.

\item \textit{Pureness Index} 
($\phi \leq 1$) instead measures the ability to
identify all edge-nodes. When $\phi = 1$, DBSCAN assigns each GT-label to at least one cluster, i.e., it correctly identifies all \nodes. $\phi < 1$ indicates that some \nodes disappear into other clusters (i.e., their GT-label is not the majority label for any cluster). Third sub-figure of Fig.~\ref{fig:indexes} reports an example where the clustering algorithm groups together \nodes with GT-labels E-2 and E-3 in cluster C-2, thus leading to $\phi < 1$.

\end{enumerate}

Rightmost sub-figure in Fig.~\ref{fig:indexes} also depicts the ideal clustering result in which DBSCAN groups correctly the caches for all the \nodes, i.e., one cluster for each GT-label (\node), leading to the case in which all the clustering performance indices, $TPR$, $\mu$ and $\phi$, are equal to 1.

Finally, we use also the number of noise points as an index of bad clustering results, i.e., the inability of DBSCAN to group caches into \nodes.

\begin{figure}[t!]
\subfigure[{DBSCAN with percentiles as features.}]{
    \includegraphics[width=1.0\columnwidth]{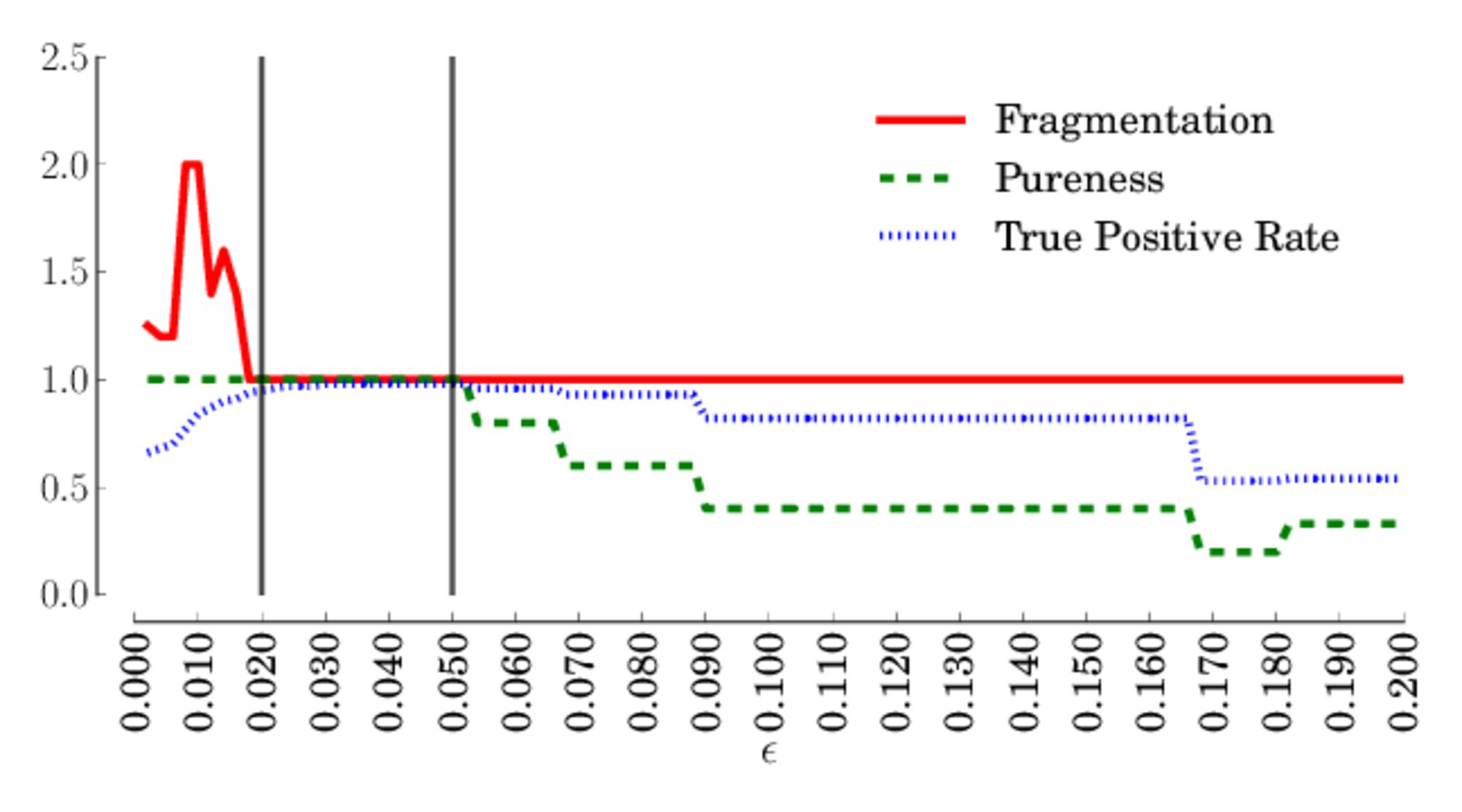}
    \label{fig:performance_perc}}
\subfigure[{DBSCAN with mean and standard deviation as features.}]{
    \includegraphics[width=1.0\columnwidth]{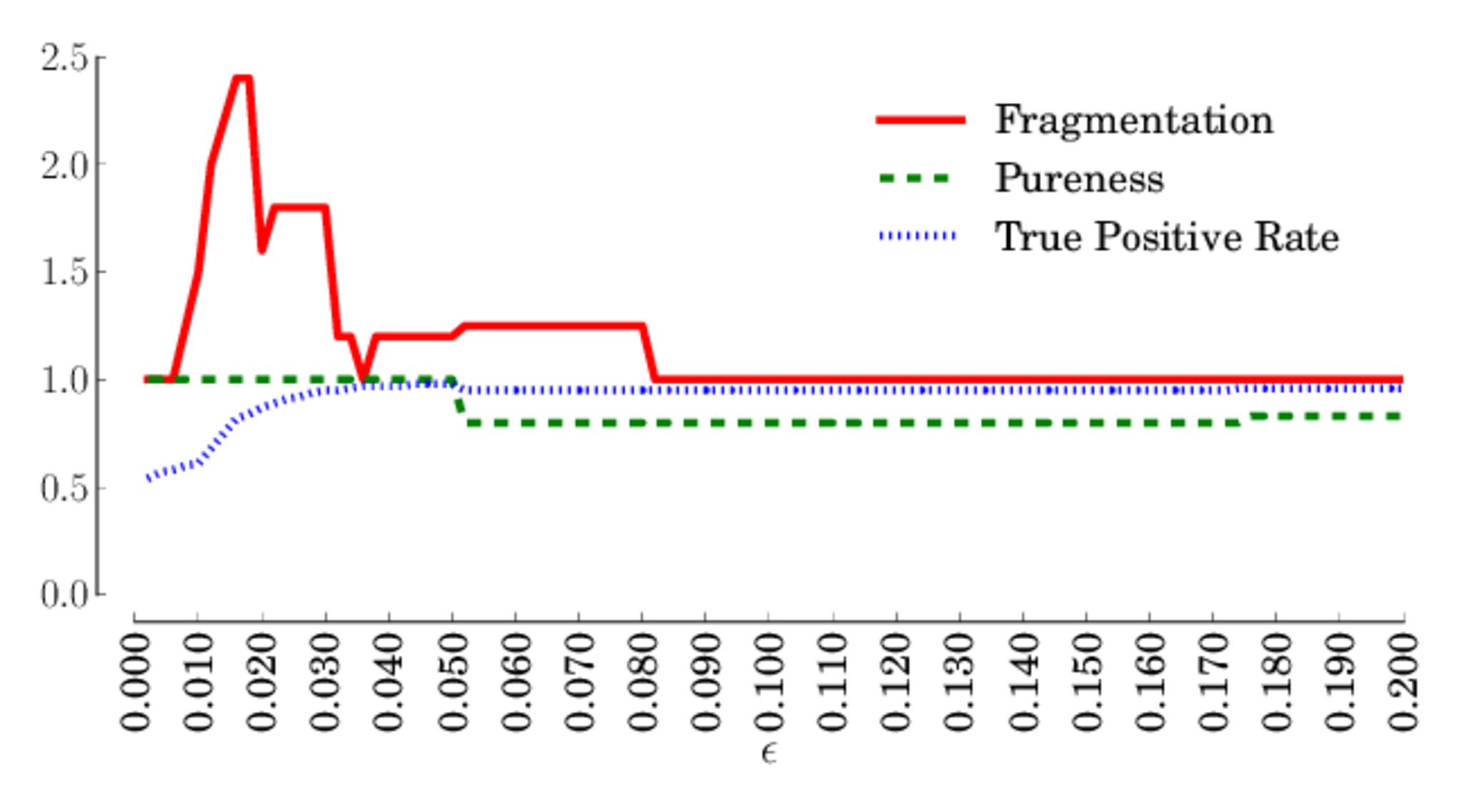}
    \label{fig:performance_std}
    }
\caption{DBSCAN with different feature settings. Performance versus $\epsilon$. 1st week of November, \TApub.}
\end{figure}

\subsubsection{DBSCAN Performance and Parameter Sensitivity}

We run experiments to evaluate the impact of DBSCAN parameters, i.e., the choice of the features, $MinPts$ and $\epsilon$.
For now, we set features as the 20th, 35th, 50th, 65th, 80th percentiles for both the RTT and TTL distributions.
$MinPts$ is typically not critical since it defines the minimum number of caches in an \node DBSCAN needs to form a cluster. We set it to 5.
Instead, we must choose $\epsilon$ carefully: If too small, a lot of fragmented clusters will emerge, or a large number of points will not be able to form dense areas, increasing the number of noise points; conversely, large values tend to create few, very large clusters, that aggregates caches from different \nodes.

Fig.~\ref{fig:performance_perc} reports the clustering indices when varying $\epsilon \in [0.0:0.2]$. As shown, we achieve the best performance with values between $0.018$ and $0.052$ (in between the vertical solid lines). For such values, all the three indices are equal or very close to 1.
Smaller values of $\epsilon$ increase the number of noise points and artificially fragment \nodes into multiple clusters. TPR decreases, while $\mu$ first increases, then decreases due to caches DBSCAN labels as noise (more than 300 caches fall in the noise for $\epsilon<0.005$).
For $\epsilon$ larger than 0.052 DBSCAN merges \nodes into too few clusters, and both $\phi$ and the $TPR$ considerably decrease.

We repeat this analysis for other traces and for different snapshots. We find $\epsilon \in [0.02:0.045]$ to give consistent results. In the following we choose $\epsilon=0.04$.

We also run a set of experiments to choose which features to use to capture the RTT and TTL distributions.
We replace the vector of percentiles $P_m(x)$ in Eq.(\ref{eq:norm}) with simple statistics, e.g., the mean and the standard deviation. The goal of this experiment is to verify whether we can replace the percentiles with some measure which does not require us to build an empirical distribution, a task which requires to collect a fairly large number of flows per cache.

Fig.~\ref{fig:performance_std} depicts results for varying $\epsilon$. Unfortunately, DBSCAN shows a good clustering for a tiny interval of values of $\epsilon$, e.g., $\epsilon=0.035$. For $\epsilon>0.035$, DBSCAN merges \nodes together, so that $\mu>1$ and $\phi<1$. By investigating further, we observe that the mean and standard deviation vary widely among caches in the same \node. This variability is due to the tails of the distributions which include outliers, e.g., very large RTT samples which bias the mean and standard deviation, but have little or no impact on the percentiles. Indeed, the percentiles of caches in the same \node are very similar, except those that gauge the tail (see the 95th percentiles in Fig.\ref{fig:perc1}).
This suggests that the choice of the percentiles to populate the vector $P_m(x)$ is more robust with respect to other simpler statistics. We run other experiments with different percentile choices that we do not report for the sake of brevity. We observe no significant differences if we avoid considering percentiles in the tail. Similarly, we observe that using both RTT and TTL gives better results than considering RTT or TTL alone.

\section{\toolintitle's highlighting capability}
\label{sec:highlight}

\begin{figure}[t!]
\begin{center}
\subfigure[{\TApri}]{
 \centering
 \includegraphics[width=1.0\columnwidth]{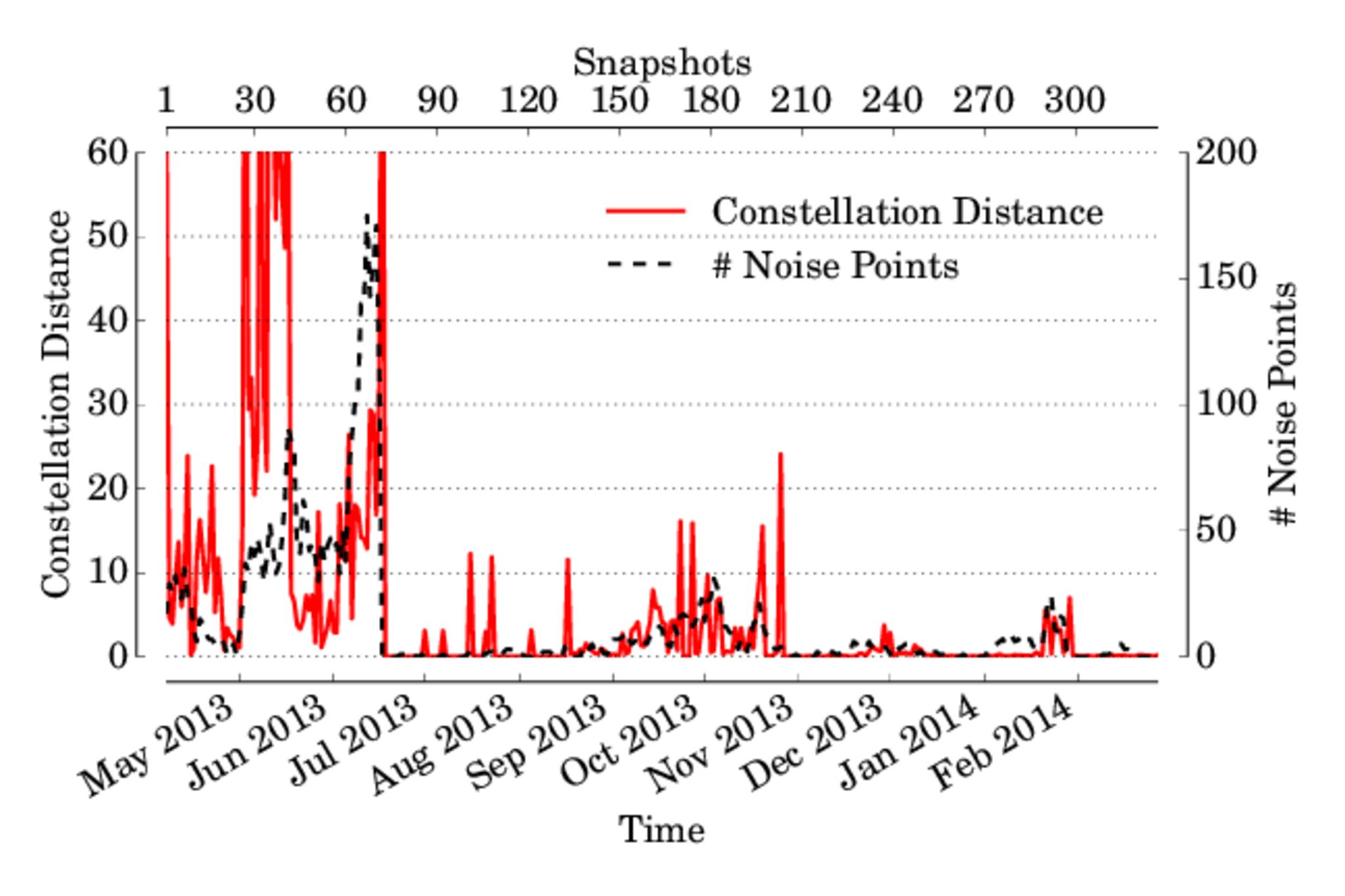}
 \label{fig:costelation_values-2}
}
\subfigure[{\TApub}]{
 \centering
 \includegraphics[width=1.0\columnwidth]{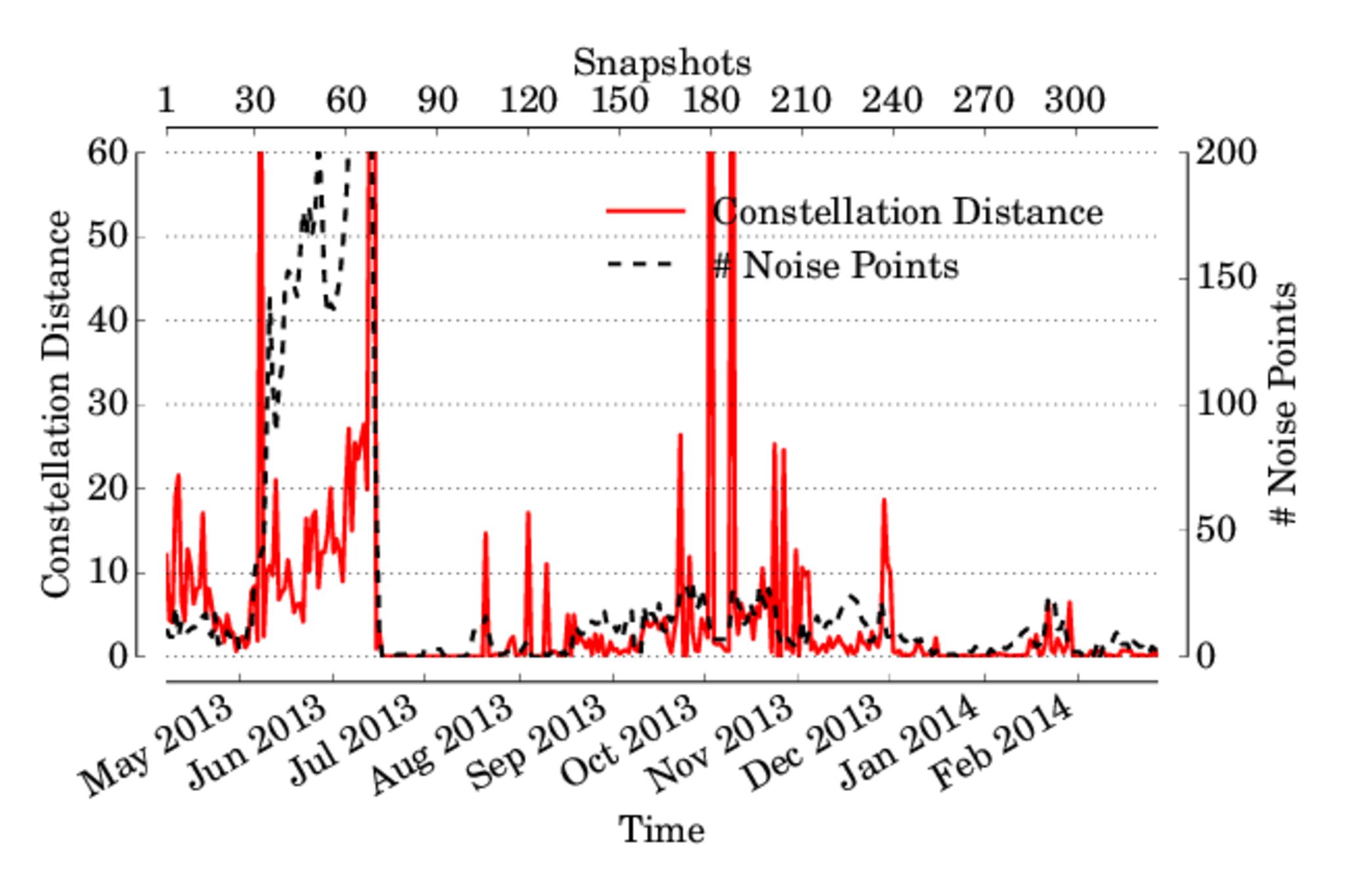}
 \label{fig:costelation_values-1}
}
\subfigure[{\TBpub}]{
 \centering
 \includegraphics[width=1.0\columnwidth]{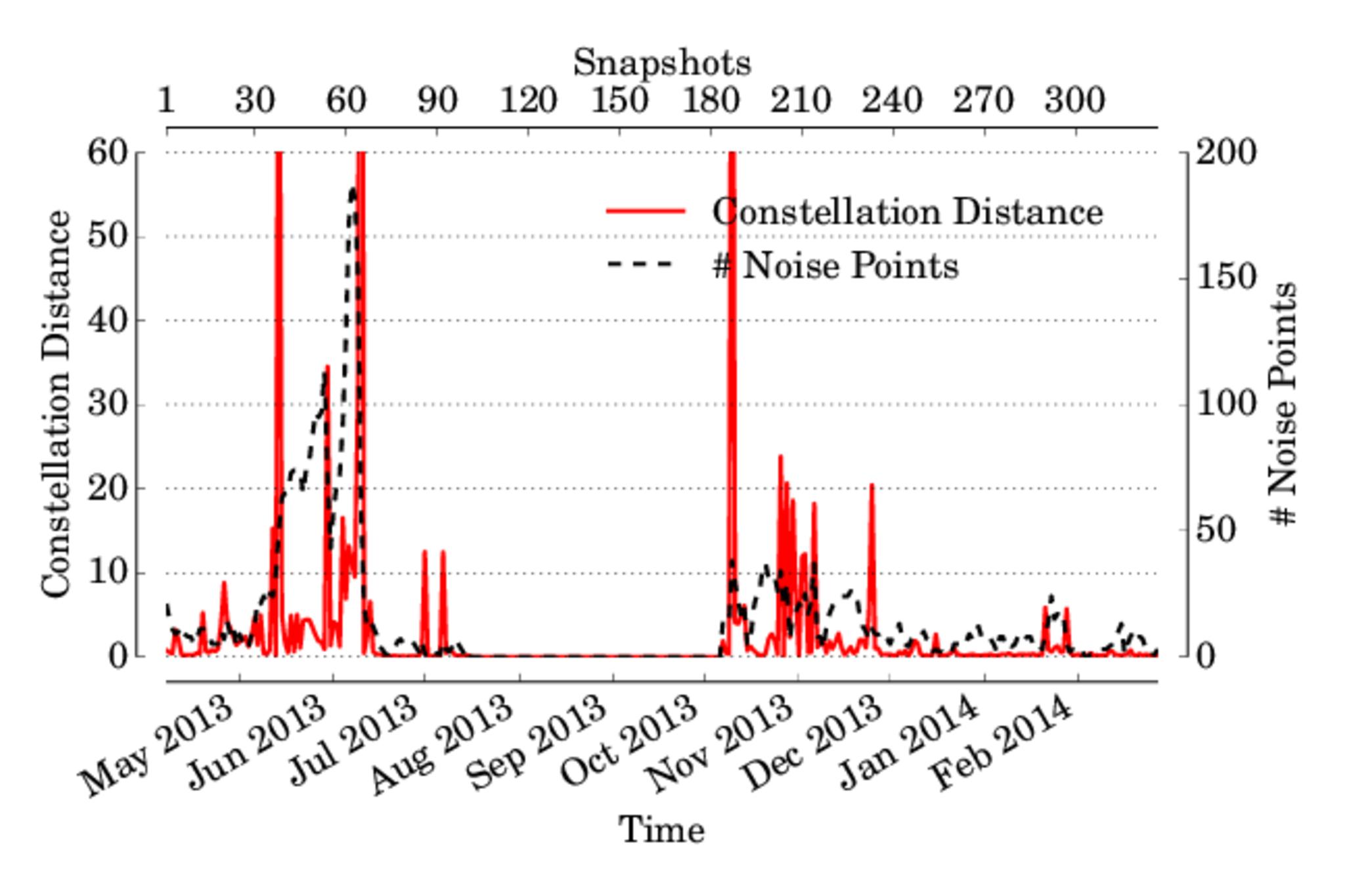}
 \label{fig:costelation_values-3}
}
\caption{\distance values and number of noise points for different traces from ISP1.}
\label{fig:costelation_values}
\end{center}
\end{figure}

In this section we run \tool over the four traces in Tab.~\ref{tab:desc-traces} to validate its capability of highlighting changes in the YouTube CDN.
The rationale is to let the ISP observe macroscopic changes that may affect a large number of users, and which may last for moderate time periods.
We consider $\Delta T=7$~days, and we start a new snapshot at midnight of every day. Snapshots form a sliding window that moves forward every day, and aggregates statistics for the past seven days. $\Delta T=7$~days guarantees to collect large enough number of samples for the large subset of the most used caches.

Fig.~\ref{fig:costelation_values} shows the evolution of the \distance (red solid curve, left y-axes) over time. It also depicts the evolution over time of the number of caches that remain in the noise after clustering (black dashed curve, right y-axes). From top to bottom, plots refer to \TApri, \TApub and \TBpub. X-axes reports daily snapshots, starting from April 1st, 2013.\footnote{PoP referring to \TBpub suffered an outage from mid July 2013 to the end of September.}

As shown, the \distance is very good at highlighting events. 
Indeed, according to Sec.~\ref{sec:observation}, a $CD>10$ suggests that the clustering at time $(n)$ is very different to the one at time $(n+1)$. Thanks to the data aggregation we obtain with the clustering, we can easily analyze the highlighted events, and quickly identify the edge-nodes involved in the changes. We investigate these events, and verify that they all correspond to sudden changes in the \nodes used by YouTube in serving ISP customers. In the following, we illustrate the most relevant ones, i.e., those with a $CD>50$.


\subsection{Large event, involving all ISP customers}
We first investigate an event \tool highlights in three different datasets. It starts on May 2nd (snapshot 27), May 7th (snapshot 32), and May 13th (snapshot 38) for \TApri, \TApub and \TBpub, respectively. \distance peaks above 60. Starting from then, both $CD$ and the number of noise points are very large. This indicates an unstable behavior, with many caches that DBSCAN cannot successfully group together, and the clustering pattern that keeps changing day by day, for more than 40 days.

\begin{figure}[t!]
\centering
    \includegraphics[width=1.0\columnwidth]{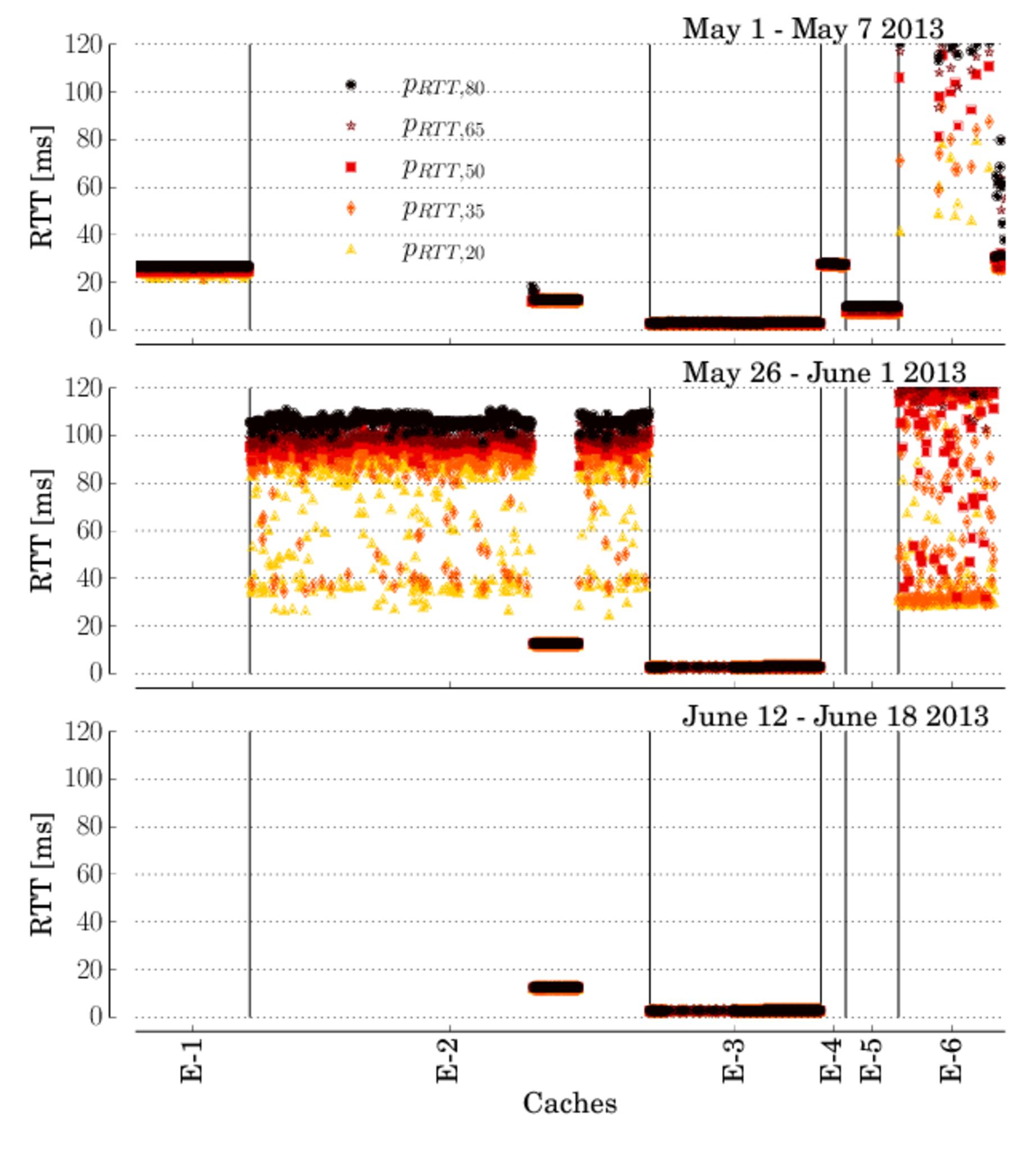}
    \caption{Per-cache RTT percentiles during the ISP-wide anomaly in May 2013. Dataset \TApub.}
    \label{fig:percAnom}
\end{figure}


To give the intuition of what happened, Fig.~\ref{fig:percAnom} shows the per-cache percentiles of the RTT that we measure in \TApub before, during, and after the anomalous event. 
First, we notice that most of the \nodes suddenly change: \AMS, \SWI, \SWII and \LND actually ``disappear'' from the clustering pattern, and during the event, many previously unseen caches in \node \FRA start serving lots of customers (observe the center plot). Second, and more surprisingly, the path properties to these new caches is by far different from paths to other caches in \FRA: the RTT percentiles are much larger ($95$ms versus $15$ms for the 50th percentile) and much more variable.
Despite these caches share the same IATA code (\FRA), the path to reach them is different from the path of other caches in \FRA, with the former possibly being severely congested. Some of these caches form new clusters, but most of them become part of the noise: Indeed, their features do not correspond to the ones DBSCAN's tuning is expecting, i.e., the distance between points is higher then $\epsilon=0.04$. We call these caches Bad-\FRA, in opposition to the small share of caches still belonging to \FRA, but showing small RTT, i.e., Good-\FRA.

\begin{figure}[t!]
\centering
    \includegraphics[width=1.0\columnwidth]{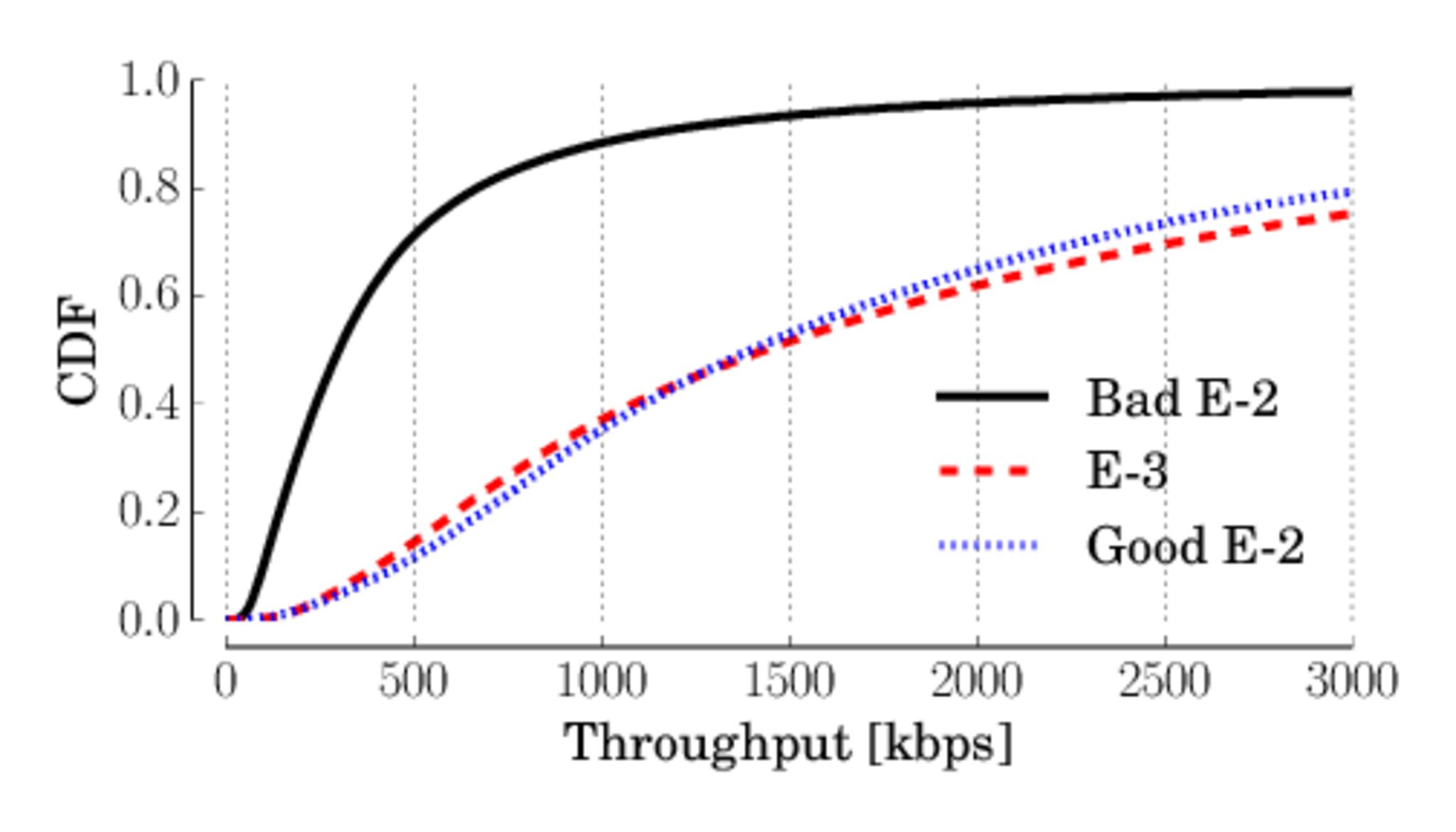}
    \caption{Throughput distribution for flows served by \MIL, Good-\FRA and Bad-\FRA during the large anomaly we observe in May 2013. Dataset \TApri.}
    \label{fig:thru}
\end{figure}

We now analyze the impact of such change on the Quality of Experience the ISP customers perceive. We report in Fig.~\ref{fig:thru} the distributions of the download throughput obtained by video retrieved by caches in \MIL, the best \node to ISP customers, Good-\FRA and Bad-\FRA. The difference is striking: while videos served by \MIL and Good-\FRA have throughput that allows to enjoy YouTube with no major impact on the QoE ($>$1,000~kb/s in 63\% of the cases), the throughput for Bad-\FRA caches is below 500~kb/s (250~kb/s) in 75\% (40\%) of the cases, clearly not enough to enjoy a video with a satisfiable QoE.
Tab.~\ref{tab:formats} corroborates above observation reporting the fractions of video (and audio) formats seen in flows handled by both Good-\FRA and Bad-\FRA. For this analysis we consider only DASH formats, as for these formats the cache delivering the video automatically adapts the quality of the video stream depending on the congestion it measures on the path to the client.  As shown,  Good-\FRA serves larger fractions of high-definition videos. Conversely,  the share of videos encoded with low-definition (144p and 240p) increases for Bad-\FRA. This confirms that Bad-\FRA experienced possible congestion during the monitored period, severely impairing the QoE of the users.

\begin{table}[t!]
\centering
\resizebox{0.60\columnwidth}{!}{\centering
\begin{tabular}{c|r|r}
\textbf{Format} & \textbf{Good-E2} & \textbf{Bad-E2} \cr
\hline
144p        & 17.4\%      & 31.7\% \cr
240p        & 18.3\%      & 26.1\% \cr
360p    & 45.4\%      & 35.7\% \cr
480p        & 14.5\%      &  5.3\% \cr
720p        & 3.8\%       &  1.0\% \cr
1080p       & 0.6\%       & 0.2\% \cr
\hline
AAC128 & 80.3\% & 92.0\% \cr 
AAC256 & 19.7\% & 8.0\% \cr 
  \end{tabular}}
  \caption{Fractions of video and audio DASH formats served by Good-\FRA and Bad-\FRA. Dataset \TApri.}
  \label{tab:formats}
\end{table}

By double checking this event with the ISP network support team, we confirm the incident involved most of their customers, increasing dramatically the complaining at their customer support.
This confirms the pervasiveness of this event upon ISP customers.


\subsection{Other events for ISP1}
We manually cross check other events, and find that some of those affected only part of the ISP customers.
This shows that YouTube CDN allocates customers to \nodes using a fine grained granularity, i.e., the 
load-balancing allows to identify small groups of clients by using the client IP address (or network).
For instance, on October 2nd (snapshot 180) and October 9th (snapshot 187) \tool highlights two sudden changes in the \TApub and \TBpub, as the \distance peaks over $60$.
Inspecting the astral distances one by one, we observe that the changes are due to 3 \nodes(\SWI, \SWII and \LND) out of 7 that suddenly ``appear'' in snapshot 180 and ``disappear'' in snapshot 187. The remaining four \nodes then serve the videos for customers in \TApub and \TBpub. We analyze the impact of the presence of such caches on the QoE by measuring the aggregate download throughput before, during and after their permanence, but we do not appreciate any significant change. Also in this case we double check the event with the ISP support team and we confirm that the change had no influence on the QoE as the customer support did not receive any meaningful complaining in the considered period.

Finally, observe that for \TApri, we do not detect any change ($CD=0.12$) in the same period, as YouTube's CDN keeps serving customers with the same group of \nodes, and we do not notice any impact on the QoE for this event too.

\subsection{Events in ISP2}

\begin{figure}[t!]
 \centering
 \includegraphics[width=1.0\columnwidth]{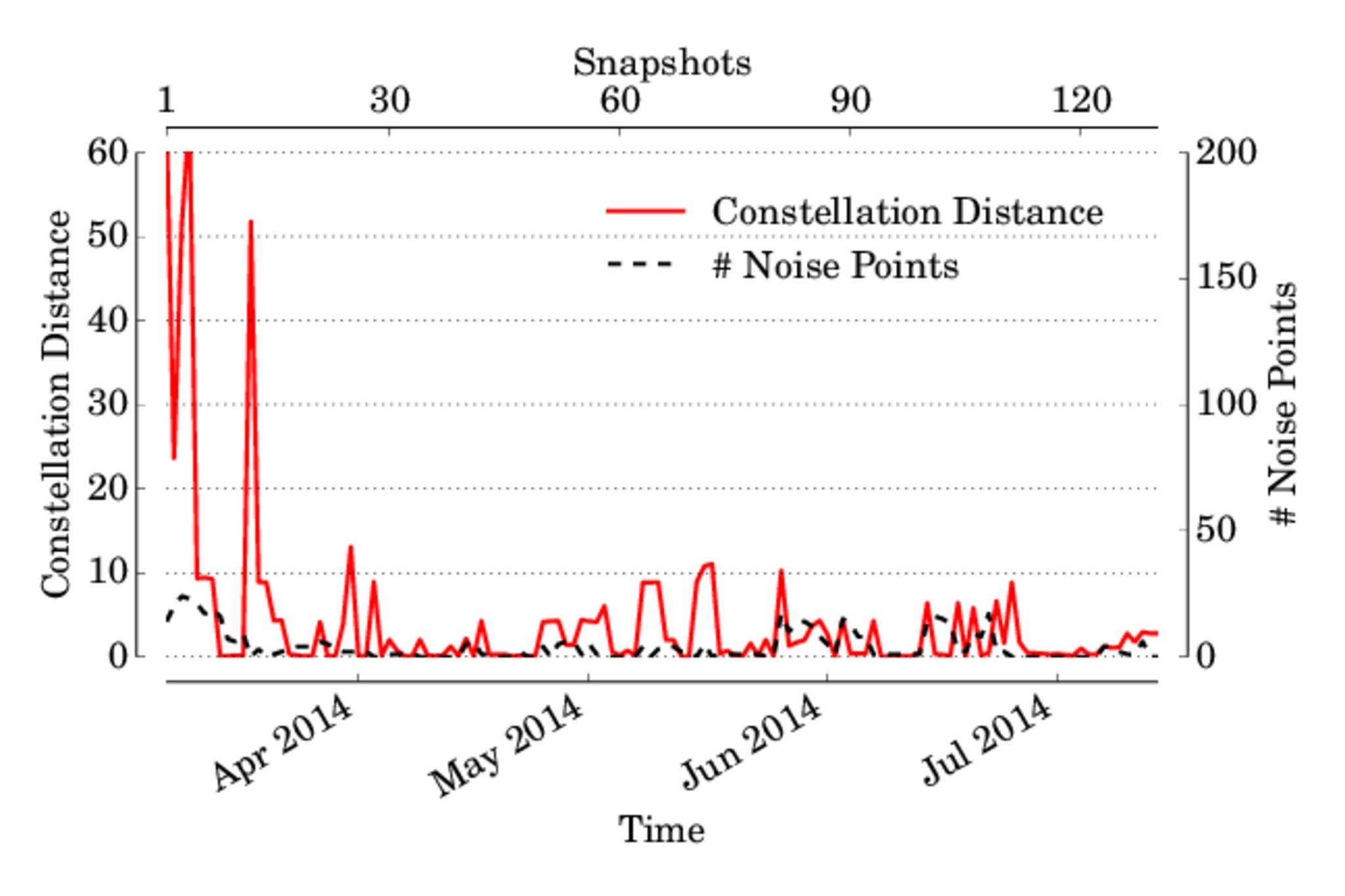}
 \caption{\distance values and number of noise points for dataset \TC.}
\label{fig:costelation_polonia}
\end{figure}

As a last set of experiments, we run \tool on the \TC dataset, which we recall we collect in ISP2, a different ISP in a different country. We run \tool with the same parameters we tune for ISP1, i.e., without going through $\epsilon$ optimization.
Indeed we aim to check whether if the \node model that DBSCAN creates is general and robust enough to work in a completely different scenario.

\begin{figure}[t!]
\centering
    \includegraphics[width=1.0\columnwidth]{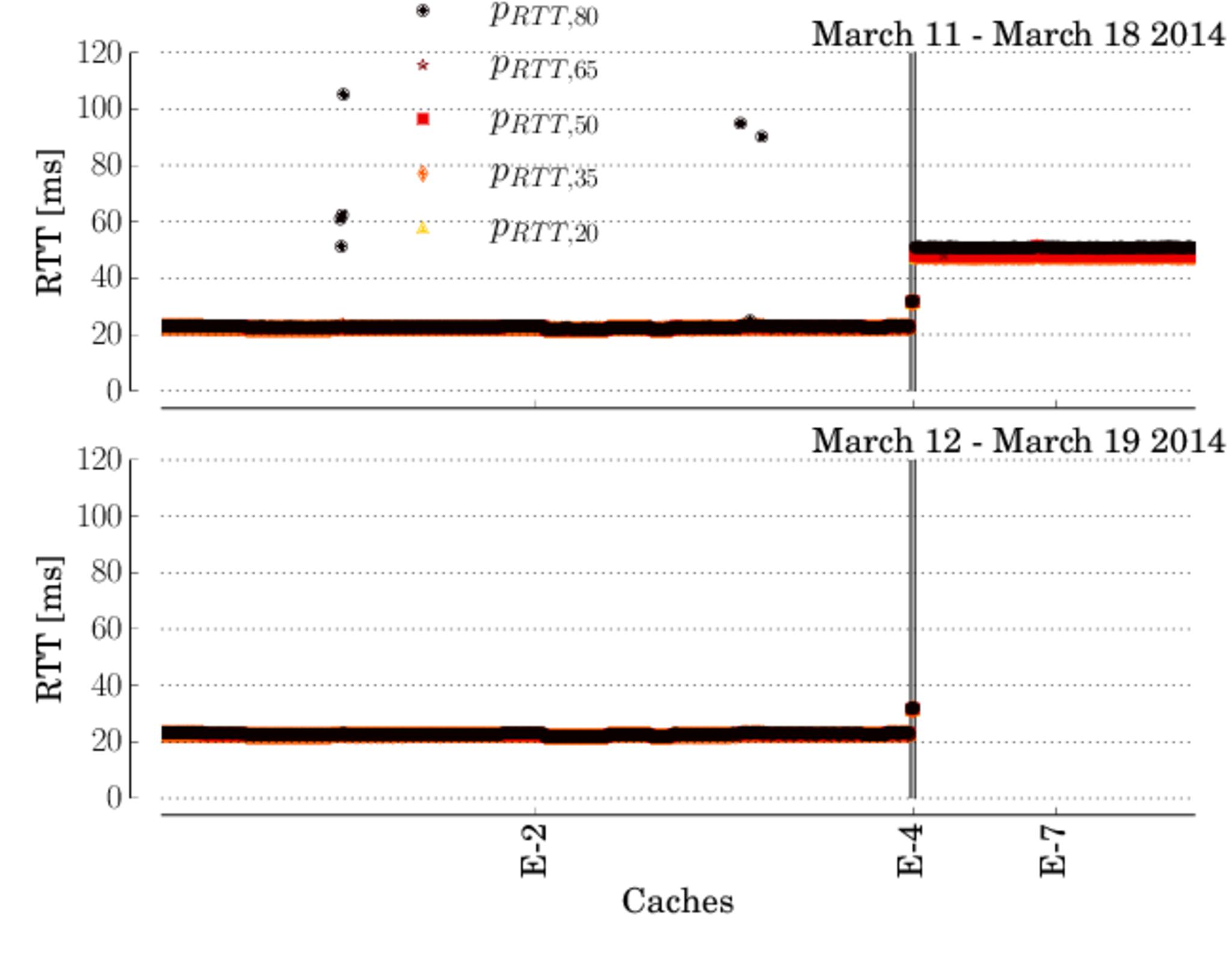}
    \caption{Per-cache RTT percentiles during the second ISP-wide anomaly in March 2014. Dataset \TC.}
    \label{fig:CD_poland}
\end{figure}

We repeat the experiment of Fig.~\ref{fig:costelation_values} for \TC dataset, and we analyze the evolution of the \distance and number of noise points. We report the results in Fig.~\ref{fig:costelation_polonia}.
To check if the clustering correctly identifies the \nodes, we select five different snapshots at random among the ones where \tool highlights no events. Again, we use the IATA codes as ground truth, and we manually check IP address subnets, RTTs and TTLs to see if some suspicious cache is present in a cluster. The clustering results in perfect match with the (possible) \nodes in the ground truth.
This despite \nodes, path, and ISP in this dataset are completely different.

We then check two suspicious events. The first one occurs from March 7th to March 10th, 2014 (snapshots 1-4, $CD>60$), and the second one happens on March 18th, 2014 (snapshot 12, $CD>51$). We observe that the first anomaly is due to a change in the network path to reach a small group of caches in \FRA. We observe that this deviation does not influence the QoE perceived by the users. For the second event, by comparing the clustering at snapshot 12 with the following snapshot, i.e., snapshot 13 (March 19th), we observe a notable change in the infrastructure of the YouTube CDN: as depicted in Fig.~\ref{fig:CD_poland} which compares the per-cache RTT percentiles, all caches belonging to \node \MAD disappear. Also in this case, the change has no evident impact on users' QoE, as the average download throughput does not vary. However, we notice that the \node \MAD represents a much more expensive route for the ISP2, since it is located in an remote ISP for which no peering agreements are in place.

%
%
%
%

\section{Conclusions}
\label{sec:conclu}

In this paper we proposed a novel system, named \tool, that leverages passive observation of network traffic and unsupervised machine learning techniques to automatically monitor and identify changes in the YouTube CDN. Based on the well known DBSCAN clustering algorithm, \tool is able to automatically group thousands of caches into few \nodes. 
To then compare the results of clustering obtained considering different snapshots collected in consecutive time intervals, we propose the \distance, a novel framework that, for the first time to the best of our knowledge, allows to easily pinpoint changes in clusters.

\tool is validated using a large dataset of traces reporting the activity of users regularly accessing YouTube. Our results are excellent: we show that after a short and simple tuning procedure to find the best setup for DBSCAN, \tool can detect anomalous events that happened in YouTube CDN. For instance, we could notice a large transformation in a crucial \node of YouTube CDN which notably impaired the QoE perceived by the monitored ISP customers for more than 40 days. 

We believe that \tool may represent a promising opportunity for ISPs, network administrators, developers and researchers to monitor the traffic generated by YouTube CDN. ISPs, for instance, may employ \tool to design automatic traffic engineering policies or to promptly react when changes in YouTube CDN impair the QoE of their customers.

Our ongoing efforts are focused on three directions: First, we are working to automate the tuning of \tool's parameters, and, thus, its whole operation process. Second, we are developing an online deployment of \tool, capable of detecting changes in YouTube CDN in real time. Third, we are adapting it to consider other use cases.

%
%

%
%
\bibliographystyle{abbrv_ccr}

%
%

\end{document}